\documentclass[useAMS,usenatbib]{mn2e}

\usepackage{times}

\usepackage{graphicx}
\usepackage{times}
\newcommand{\cthead}[1]{\multicolumn{1}{c}{#1}}
\newcommand{\kss}{km~s$^{-1}$ }
\newcommand{\ks}{km~s$^{-1}$}

\title[Methanol masers near IRAS 16547-4247]
{Class I methanol masers in the outflow of IRAS 16547-4247}
\author[M. A. Voronkov et al.]{M. A. Voronkov$^{1,2}$\thanks{E-mail:
Maxim.Voronkov@csiro.au}, K.J. Brooks$^{1}$, A.M. Sobolev$^3$,
S.P. Ellingsen$^{4}$, \newauthor
A.B. Ostrovskii$^3$ and J.L. Caswell$^1$\\
$^{1}$Australia Telescope National Facility CSIRO, PO Box 76, Epping, NSW 1710, Australia\\
$^{2}$Astro Space Centre, Profsouznaya st. 84/32, 117997 Moscow, Russia\\
$^{3}$Ural State University, Lenin ave. 51, 620083 Ekaterinburg, Russia\\
$^{4}$School of Mathematics and Physics, University of Tasmania, GPO Box
252-37, Hobart, Tasmania 7000, Australia\\}
\begin{document}

\date{}

\pagerange{\pageref{firstpage}--\pageref{lastpage}} \pubyear{2005}

\maketitle

\label{firstpage}

\begin{abstract}
  The Australia Telescope Compact Array (ATCA) has been used to image
  class I methanol masers at 9.9, 25 (a series from J=2 to J=9), 84,
  95 and 104~GHz located in the vicinity of IRAS 16547-4247
  (G343.12-0.06), a luminous young stellar object known to harbour a
  radio jet.  The detected maser emission consists of a cluster of 6
  spots spread over an area of 30\arcsec. Five spots were detected in
  only the 84- and 95-GHz transitions (for two spots the 84-GHz
  detection is marginal), while the sixth spot shows activity in all
  12 observed transitions. We report the first interferometric
  observations of the rare 9.9- and 104-GHz masers. It is shown that
  the spectra contain a very narrow spike ($<0.03$~\ks) and the
  brightness temperature in these two transitions exceeds
  $5.3\times10^7$ and $2.0\times10^4$~K, respectively. The three most
  southern maser spots show a clear association with the shocked gas
  traced by the H$_2$ 2.12~$\mu$m emission associated with the radio
  jet and their velocities are close to that of the molecular core
  within which the jet is embedded. This fact supports the idea that
  the class~I masers reside in the interface regions of
  outflows. Comparison with OH masers and infrared data reveals a
  potential discrepancy in the expected evolutionary state.  The
  presence of the OH masers usually means that the source is evolved,
  but the infrared data suggest otherwise. The lack of any class~II
  methanol maser emission at 6.7~GHz in the source raises an
  additional question, is this source too young or too old to have
  a 6.7~GHz maser? We argue that both cases are possible and suggest
  that the evolutionary stage where the class I masers are active, may
  last longer and start earlier than when the class II masers are
  active. However, it is currently not possible to reveal the exact
  evolutionary status of IRAS 16547-4247.
\end{abstract}


\section{Introduction}

Methanol masers are commonly found in massive star-forming
regions. They fall into 2 categories first defined by \citet{bat88}
and further considered by \citet{vor05b}: class~II masers (of which
the 6.7 GHz is the best known and usually strongest) are closely
associated with infrared sources  
and reside in the close environment of exciting (proto-)stars. In
contrast, class~I masers (e.g. 95 GHz) are usually found offset (by up
to a parsec) from continuum sources
\citep*[e.g.,][]{kur04}. Theoretical calculations strongly suggest
that class~I masers are pumped via collisions with the molecular
hydrogen in contrast to class II masers, which have a radiative pump
\citep[e.g.,][]{cra92,vor99,vor05b}.

Class I masers are relatively poorly studied. There is direct
observational evidence to suggest a relationship between these masers
and outflows in a number of sources \citep{pla90,men96}, although
another scenario, which involves cloud-cloud collisions, has also been
proposed \citep*{sob92,meh96,sal02}. The common point of these two
scenarios is the presence of shocks. There is observational evidence
that methanol abundance is significantly increased in the shock
processed regions \citep[e.g.,][]{gib98} and the gas is expected to be
heated and compressed in such regions providing more frequent
collisions and, therefore, more efficient pumping.

\begin{table*}
\caption{Dates of observations, array configurations and observed transitions}
\label{obstab}
\begin{tabular}{@{}r@{}c@{}r@{ }r@{}c@{}c@{}crrc}
\hline
\cthead{UT}&\cthead{Array}&
\cthead{Transition}&\cthead{Frequency}&\cthead{Velocity}&\cthead{Primary}&\multicolumn{2}{c}{Synthesized beam}&
\cthead{Spectral}&\cthead{Velocity coverage}\\
\cthead{Date}&\cthead{configuration}&
\multicolumn{2}{c}{}&\cthead{Uncertainty}&\cthead{beam}&\cthead{FWHM}&\cthead{pa}&
\cthead{Resolution}&\\
&&&\cthead{(MHz)}&\cthead{(\ks)}&\cthead{(arcmin)}&
\cthead{(arcsec)}&\cthead{(\degr)}&
\cthead{(\ks)}&\cthead{(\ks)}\\
\hline
 2 May 2005 & 750A &$\hphantom{1}9_{-1}-\hphantom{1}8_{-2}$~E\hphantom{$^+$}& \hphantom{2}9936.202\hphantom{0}~(4)& 0.12\hphantom{0} &5.09 & \multicolumn{2}{c}{no imaging}& 0.12\hphantom{7}& $-$92, $+$28\\
            &      &$\hphantom{1}2_{2\hphantom{-}}-\hphantom{1}2_{1\hphantom{-}}$~E\hphantom{$^+$} & 24934.382\hphantom{0}~(5) & 0.06\hphantom{0} & 2.03 & 2.0$\times$1.3 & 88 & 0.047 & $-$36, $+$41\\
            &      &$\hphantom{1}3_{2\hphantom{-}}-\hphantom{1}3_{1\hphantom{-}}$~E\hphantom{$^+$} & 24928.707\hphantom{0}~(7) & 0.08\hphantom{0} & 2.03 & 1.2$\times$0.8 & 6 & 0.047 & $-$104, $-$27.5\\
            &      &$\hphantom{1}4_{2\hphantom{-}}-\hphantom{1}4_{1\hphantom{-}}$~E\hphantom{$^+$} & 24933.468\hphantom{0}~(2) & 0.02\hphantom{0} & 2.03 & 1.2$\times$0.8 & 14 & 0.047 & $-$46, $+$30\\
	    &      &$\hphantom{1}5_{2\hphantom{-}}-\hphantom{1}5_{1\hphantom{-}}$~E\hphantom{$^+$} &24959.0789~(4)& 0.005 & 2.03 & 1.2$\times$0.7 & 22 & 0.047 & $-$55, $-$18\\
	    &      &$\hphantom{1}6_{2\hphantom{-}}-\hphantom{1}6_{1\hphantom{-}}$~E\hphantom{$^+$} &25018.1225~(4)& 0.005 & 2.02 & 1.6$\times$0.8 & 20 & 0.047 & $-$54, $-$18\\
16 Jun 2005& 6B    &$\hphantom{1}9_{-1}-\hphantom{1}8_{-2}$~E\hphantom{$^+$}& \hphantom{2}9936.202\hphantom{0}~(4)& 0.12\hphantom{0} & 5.09 & 1.6$\times$1.2& $-$1 & 0.029  & $-$70, $+$26\\
            &      &$\hphantom{1}6_{2\hphantom{-}}-\hphantom{1}6_{1\hphantom{-}}$~E\hphantom{$^+$} &25018.1225~(4)& 0.005 & 2.02 & 0.6$\times$0.5 & $-$5 & 0.023 & $-$51, $-$13\\
	    &      &$\hphantom{1}7_{2\hphantom{-}}-\hphantom{1}7_{1\hphantom{-}}$~E\hphantom{$^+$} &25124.8719~(4)& 0.005 & 2.01 & 0.7$\times$0.5 & $-$3 & 0.023 & $-$54, $-$16\\
	    &      &$\hphantom{1}8_{2\hphantom{-}}-\hphantom{1}8_{1\hphantom{-}}$~E\hphantom{$^+$} &25294.4165~(4)& 0.005 & 2.00 & 0.7$\times$0.5 & $-$3 & 0.023 & $-$47, $-$10\\
	    &      &$\hphantom{1}9_{2\hphantom{-}}-\hphantom{1}9_{1\hphantom{-}}$~E\hphantom{$^+$} &25541.3979~(4)& 0.005 & 1.98 & 0.7$\times$0.5 & $-$4 & 0.023 & $-$47, $-$10\\
18 Aug 2005&H214C  &$\hphantom{1}6_{2\hphantom{-}}-\hphantom{1}6_{1\hphantom{-}}$~E\hphantom{$^+$} &25018.1225~(4)& 0.005 & 2.02 &\multicolumn{2}{c}{no imaging} & 0.023 & $-$55, $-$8\\
            &      &$\hphantom{1}5_{-1}-\hphantom{1}4_{0\hphantom{-}}$~E\hphantom{$^+$} &84521.169~(10)& 0.04\hphantom{0}  & 0.60 & 2.7$\times$1.9 & 88 & 0.028 & $-$41, $-$19\\
            &      &$\hphantom{1}8_{0\hphantom{-}}-\hphantom{1}7_{1\hphantom{-}}$~A$^+$&95169.463~(10)&  0.03\hphantom{0}  & 0.53 & 2.3$\times$1.7 & 87 & 0.025 & $-$41, $-$22\\
	    &      &$11_{-1}-10_{-2}$~E\hphantom{$^+$}&104300.414\hphantom{0}~(7)& 0.02\hphantom{0} & 0.49 & 2.1$\times$1.5 & 88 & 0.022 & $-$39, $-$22\\
\hline
\end{tabular}
\end{table*}

To further investigate the relationship of different class I maser
transitions with outflows and to provide data for theoretical studies,
we have made interferometric observations of 12 class I maser transitions
towards 
IRAS~16547-4247 (G343.12-0.06) using the Australia Telescope Compact
Array (ATCA). This is a significant fraction
\citep*[50\% according to][]{mul04}
of all known class~I maser transitions, the majority of which belong to
the J$_2-$J$_1$~E series.
We observed at least one transition from each transition series known
to exhibit class~I maser activity~\citep{vor99} and covered all known
class~I maser transitions within the frequency range of existing ATCA
receivers.
Only the J=5 transition from the J$_2-$J$_1$~E series at
25~GHz has been previously observed towards IRAS~16547-4247
\citep{vor05a}, although with no position measurement.
In this paper we present the first detection of 7
other transitions from this series as well as the first detection of the
84-GHz maser towards this source. The other 3
transitions were detected in a number of single dish and ATCA maser
line surveys
\citep[Voronkov et al. unpublished observations]{sly94,val00}.  It is
worth mentioning that this source is the only example where all these
class~I maser transitions have been detected together, including the
relatively rare 9.9 and 104 GHz masers.

IRAS~16547-4257 has recently been the focus of attention as the most
luminous young stellar object (YSO) known to harbour a radio jet.  At
a distance of 2.9 kpc (Bronfman, private communication), the YSO has a
bolometric luminosity of
$6.2\times10^4$ L$_{\odot}$, equivalent to that of a single O8 Zero
Age Main Sequence (ZAMS) star. The jet was first detected by
\citet{gar03} using the ATCA at 1.4, 2.5, 4.8 and 8.6 GHz and subsequently
observed using the VLA at 8.5 GHz and 14.9 GHz \citep{rod05}. This is
the first reported case of a radio jet associated with a young O-type
star. Located at the centre of a massive dense molecular core
(approximately 10$^3$ M$_{\odot}$), the jet is driving a highly
energetic collimated bipolar outflow that extends over 1.5~pc
\citep{bro03,gar06}.


\section{Observations}
\label{obs_section}

Observations were made using the ATCA in 2005 May-August with assorted
array configurations and various correlator modes. The details,
including the spectral resolutions, velocity coverages, 
full-width at half maximum (FWHM) of the primary beam, and 
synthesized beam parameters, are summarized in Table~\ref{obstab}.  We
adopted the rest frequencies determined by \citet{mul04}. The
measurement uncertainties are given in Table~\ref{obstab} in
parentheses after each corresponding frequency. They are expressed in
units of the least significant figure.  Corresponding uncertainties in
the radial velocity are given in the next column.

The $\mbox{J=2,3}$ and 4 transitions of the J$_2-$J$_1$~E series near
25~GHz were observed in a single 8~MHz band centred on the average
frequency of these transitions. The remaining transitions were
observed using a separate frequency setup for each individual
transition.  The correlator setup used for observations of the
$\mbox{J=5}$ and 6 transitions on May 2 allowed us to observe the
24896~MHz continuum emission with a 128~MHz bandwidth simultaneously
with the spectral line observations.  However, for the subsequent
observations we chose a higher spectral resolution rather than a
continuum channel.  With the exception of the 9.9~GHz observations on
June 16 two orthogonal linear polarizations were recorded. For June 16
only one linear polarization was recorded in order to obtain a larger
number of spectral channels and achieve a spectral resolution
comparable with that of the higher frequency transitions. It is worth
noting that \citet*{wie04} found a number of mm-wavelength class~I
methanol maser transitions to be polarized. The correlator setup used
in our study did not allow us to determine polarization properties.
However, during the data reduction both polarizations, if present,
were averaged together. Hence, all the flux densities with the
exception of that at 9.9~GHz measured on June 16 are not affected by
polarization. For the latter we examined the flux density measured
from 10 individual scans taken at various hour angles assuming that
the source was unresolved and located at the position found by the
imaging. No parallactic angle dependence was found beyond the noise
level of 0.3~Jy (1$\sigma$). This fact as well as a comparison with
the May observations (see section~\ref{tempvar_section}) performed in
the dual polarization mode leads us to conclude that polarization
effects, if present, are less than the calibration uncertainty.

\begin{figure*}
\includegraphics[width=\linewidth]{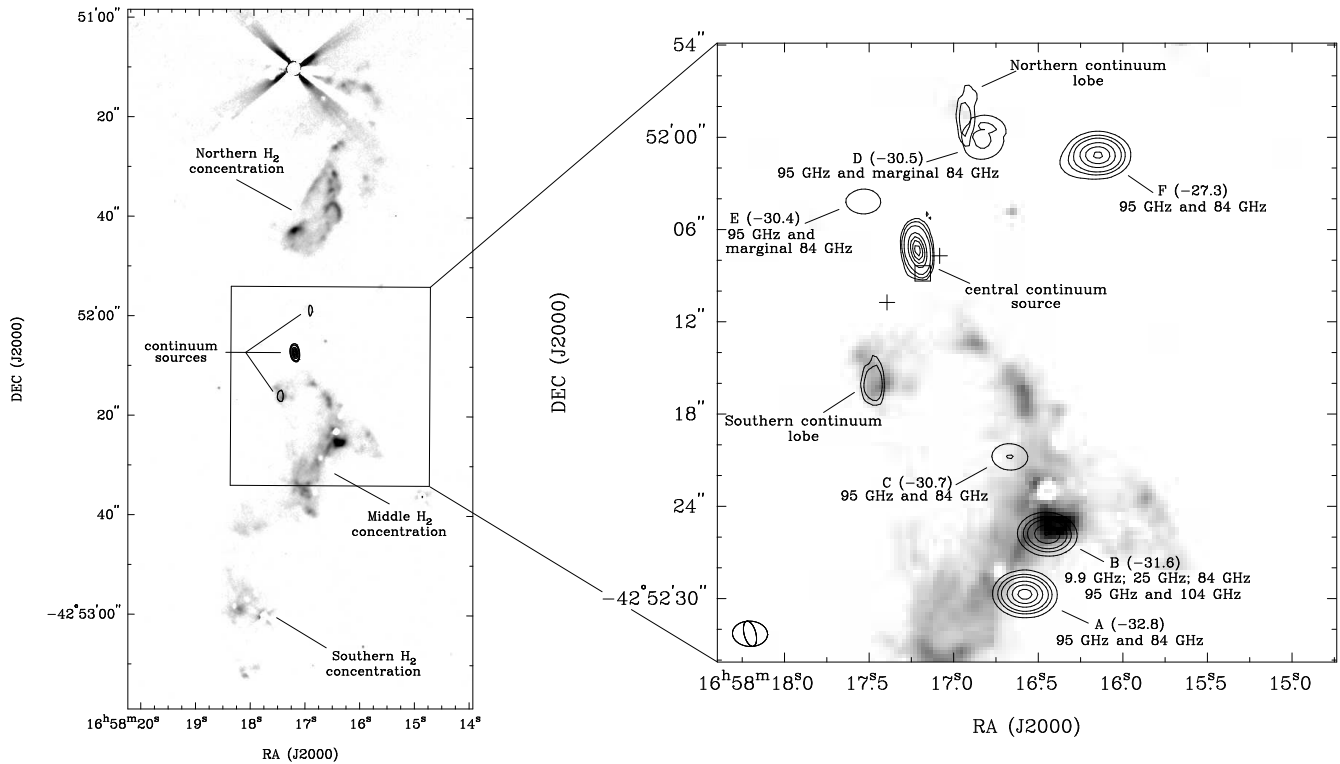}
\caption{The general morphology of the region: the 95~GHz maser spots
and the 25~GHz continuum sources are overlayed on the H$_2$ 2.12$\mu$m
image \citep{bro03}. The continuum sources are also shown on
the large scale image and believed to delineate the direction of the jet.
An artifact in the northern part of the large scale H$_2$ image is caused
by a star. Contours in the 
small scale image are (1, 2, 3, 4, 5, 7, 9)~$\times$3.6~Jy~beam$^{-1}$
and (0.3, 0.5, 1, 3, 5, 7, 9)~$\times$1.5~mJy~beam$^{-1}$
for the 95-GHz maser and 25-GHz continuum sources respectively. The lowest
contour of the 25-GHz continuum emission corresponds to 7$\sigma$. The noise
level of the 95-GHz image varies significantly across the
image (see Table~\protect\ref{spc_noise}). For spot F the lowest
contour corresponds to 9$\sigma$. The numbers
in parentheses are approximate peak radial velocities in \kss for each maser
spot. The synthesized
beam is shown in the bottom-left corner of the inset (small and large
ellipses for 25 GHz and 95 GHz, respectively). The open box and two crosses
near the central continuum source mark the positions of the mainline OH
masers and two well separated clusters of the H$_2$O masers, respectively.}
\label{genmorph}
\end{figure*}

The position of the phase and pointing centre was
$\alpha_{2000}=16^h58^m$16\fs57,
$\delta_{2000}=-42$\degr52\arcmin24\farcs0 in all observing sessions.
Reference pointing using the source 1646-50 (which served also as a
phase calibrator), was used for all observations except those at
9.9~GHz, where a global pointing model was used. From the statistics
of the pointing solution the reference pointing accuracy was estimated
to be 2\farcs3$\pm$0\farcs8 for all observing runs. The accuracy of
the global model (which affects the 9.9-GHz data only), was estimated
to be 6\arcsec$\pm$3\arcsec{ } and 10\arcsec$\pm$4\arcsec{ } for the 2
May and 16 June runs respectively.  The accuracy of the reference
pointing affects the accuracy of flux density measurements,
particularly for sources offset from the pointing centre.  The
positional accuracy of the maser locations is primarily influenced
by the quality of the phase calibration and is believed to be better
than 0.5 arcsec.  All 3mm observations were undertaken as two-point
mosaics. An additional pointing was offset half of the primary beam
width (see Table~\ref{obstab}) to the East from the position given
above.

Data reduction was performed using the {\sc miriad} package (28 August
2005 release) following standard procedures with the exception of
bandpass calibration. Using the {\sc uvlin} task of {\sc miriad} a
low-order polynomial fit to the spectrum of the bandpass calibrator
(1921-293) has been performed before attempting a bandpass solution
with the {\sc mfcal} task. This approach enables us to achieve a
higher spectral line dynamic range (i.e. detect a weaker spectral line
in front of the continuum of a given flux density), although none of
the datasets described in the paper turned out to be dynamic range
limited.

The flux density scale at 9.9 and 25~GHz was established using the
standard ATCA primary calibrator, 1934-638. For higher frequencies
planetary calibration was done using Uranus. In order to assess the
variability of the source some transitions were observed more than
once. Two such repeated observations (see Table~\ref{obstab}) were
quite short and no imaging was possible. However, an accurate check of
the flux density scale is still possible provided the source position
is the same for all epochs. The accuracy of the absolute scale of flux
density is estimated to be 3\% and 10\% at 9.9 and 25~GHz,
respectively\footnote{For details on the calibration using 1934-638 query
this calibrator at the ATCA calibrators webpage
({\it http://www.narrabri.atnf.csiro.au/calibrators})}.  The uncertainty of high frequency
planetary calibration of the ATCA has yet to be fully investigated and
may be as large as 50\%. However, based on our experience we believe it
to be better than 30\% for the calibration and data reduction
procedure we followed in our observations. This figure is in agreement
with the measured relative flux densities of the secondary calibrator
at different frequencies. An additional 8\% flux density uncertainty
related to the pointing accuracy given above exists for the high frequency
(84 and 95~GHz) features situated near the edge of the imaged field of
view.

\section{Results}
\subsection{General morphology}
\label{genmorph_section}

Figure~\ref{genmorph} shows an H$_2$ 2.12~$\mu$m image of
IRAS~16547-4247 \citep{bro03}. The image contains a complex chain of
emission with three major concentrations consistent with the
morphological characteristics of Herbig-Haro (HH) objects arising from
the interaction of a collimated flow with the ambient medium
\citep{rei01}. There are a number of mechanisms responsible for
excitation of 2.12~$\mu$m emission, the most common of which are
fluorescence and collisional excitation in shocks
\citep[e.g.][]{hab05}. Several mechanisms can simultaneously
contribute to the observed emission. In general, a detailed
spectroscopic analysis is required to determine the exact excitation
condition \citep*[e.g.,][]{eis96,wal02}.  However, the morphology of
the region, which suggests an interaction between the outflow and the
ambient medium, together with the lack of Br$\gamma$ emission which
indicates that the number of ultra-violet photons is insufficient to
pump the fluorescence \citep{bro03}, supports the idea that most of
the H$_2$ emission in this source traces shocked gas.

The outer concentrations of H$_2$ are located approximately symmetrically
offset from the brightest source of the cm-wavelength radio continuum
emission (see Fig.~\ref{genmorph}). This continuum source
was first detected by \citet{gar03} and interpreted as thermal
(free-free) emission from a radio jet. \citet{gar03} also detected two
satellite continuum sources and interpreted them as 
the internal working surfaces of the jet.
Sensitive high spatial resolution
observations carried out by \citet{rod05} resolved the northern lobe
into a chain of spots aligned along the jet direction and showed that
the peak spot most likely corresponds to optically thin thermal
emission. The southern lobe was also resolved by \citet{rod05}, but in
contrast to the northern lobe a non-thermal origin of the brightest
component has been confirmed.  Both thermal and non-thermal emission
mechanisms are expected according to theoretical studies of 
electron acceleration in shock waves \citep*[e.g.][]{hen91}.

In addition to these sources, \citet{rod05} detected three continuum
sources displaced by a few arcseconds from the jet
axis, which is thought to be a line joining the outer
continuum lobes (Fig.~\ref{genmorph}).
The brightest two of these sources are located between the southern
non-thermal lobe and the central source and form a wiggly structure,
convex in the opposite direction from the bow-shaped middle H$_2$
concentration.  \citet{rod05} proposed that one
of these continuum sources (the most distant from the jet axis) could
be associated with a young, low-mass star with a gyrosynchrotron
emission. If this interpretation is correct, the curved shape of the
middle H$_2$ concentration could be the result of an additional
outflow from this less massive star. However, it follows from the
large scale distribution of the H$_2$ emission shown in
Fig.~\ref{genmorph} that the molecular outflow covers a rather wide
area along the jet. Therefore, this morphology could also be caused by
a number of factors such as irregularities in the ambient medium or
precession of the jet \citep{bro03}.

\begin{figure*}
\includegraphics[width=\linewidth]{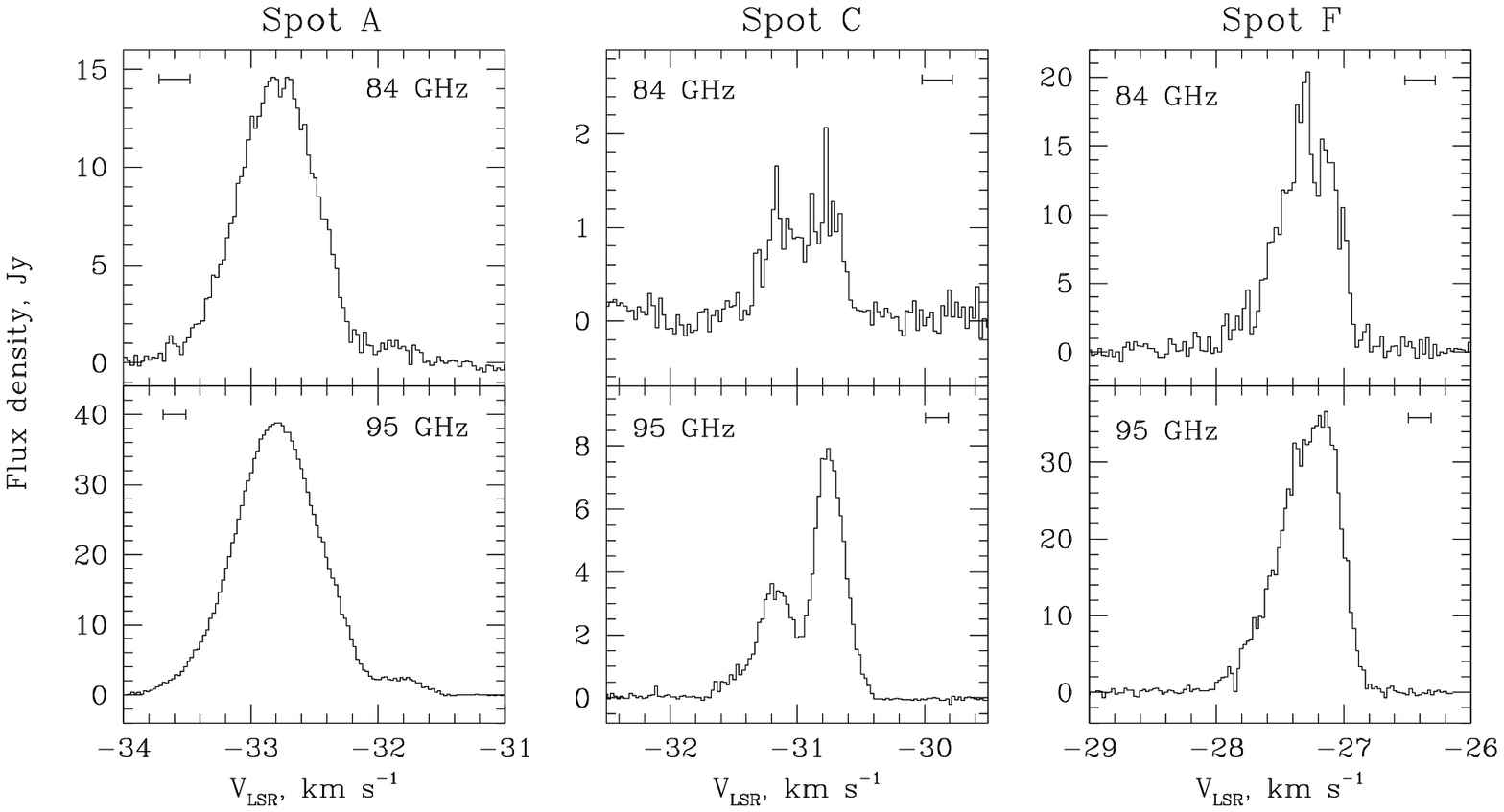}
\vskip -7mm
\includegraphics[width=\linewidth]{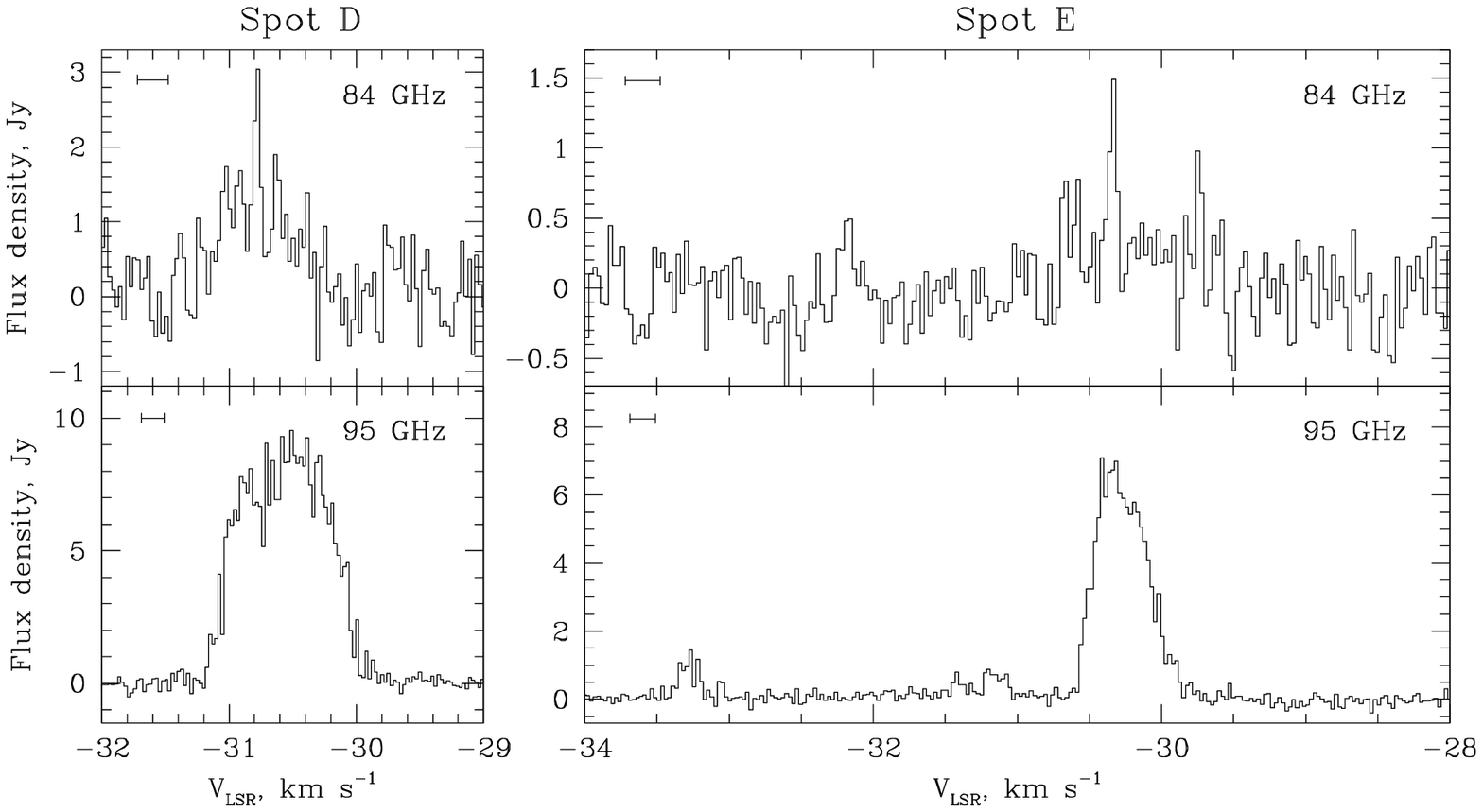}
\caption{Spectra of spots A, C -- F at 84 (upper plots)
and 95~GHz (lower plots). Spots D and E were marginal detectons at 84~GHz.
The error bar near the top of each plot shows the 3$\sigma$ velocity
uncertainty corresponding to the uncertainty in rest frequency
(see Table~\ref{obstab}).}
\label{spotAC_Fsp}
\end{figure*}

The expanded scale inset of Fig.~\ref{genmorph} shows the location of all
detected maser spots at 95~GHz (labelled A to F).  The 25~GHz
continuum emission also shown in the figure closely resembles the
triple-source morphology found at lower frequencies by
\citet{gar03}. This continuum measurement and the spectral energy
distribution (SED) of the source are discussed in a separate paper
\citep{bro06}. The outermost lobes of the H$_2$ emission are outside
the ATCA field of view at 3~mm.  The southern maser spots (labelled A
to C) show a remarkable correlation with the prominent H$_2$ feature
(Fig.~\ref{genmorph}).  Contrary to our expectations based on the
single-dish spectra of \citet{val00} obtained at five offset
pointings, no maser spots offset in the East-West direction were
detected. Instead, a group of spots (E, D and F) was found near the
edge of the field of view in the northern direction. Most likely this
discrepancy is related to the pointing model used at the Mopra radio
telescope during the observations of \citeauthor{val00}

\subsection{Methanol masers}
\begin{figure*}
\includegraphics[width=\linewidth]{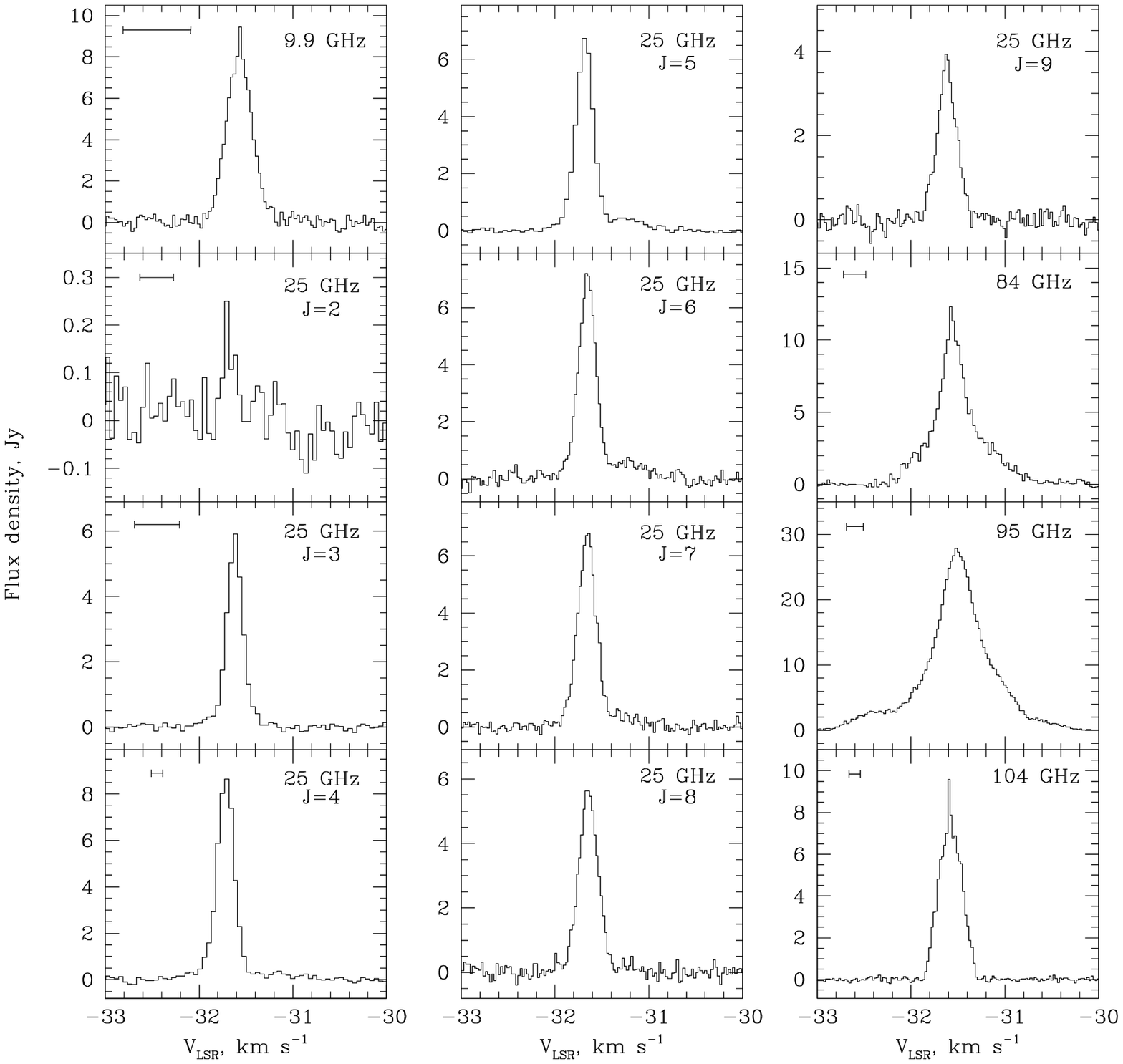}
\caption{Spectra of spot B, where masers in all observed transitions
were detected. The error bar at the top of some plots shows the 3$\sigma$
velocity uncertainty corresponding to the uncertainty in rest frequency 
(see Table~\ref{obstab}). For $\mathrm{J}=5-9$ transitions at 25~GHz this
uncertainty is below the spectral resolution and, therefore, is not shown.}
\label{spotBsp}
\end{figure*}

For each spot and observed frequency we produced spectra by summing
the flux density in each plane of the deconvolved cube after the
primary beam correction or linear mosaicing. The summation was over a
box centred on the peak pixel in the image. The size of this box was
chosen individually for each spot to suppress all unrelated emission
(which is difficult at the edge of the field of view), and was
typically about 4\arcsec.  The spectra for all spots other than spot
B, are shown in Fig.~\ref{spotAC_Fsp}.  The horizontal error bars on
these plots show the $3\sigma$ velocity uncertainty related to the
uncertainty of the rest frequency.  With the exception of spot B, all
were detected only at 84 and 95~GHz, although the 84-GHz detections for
spots D and E are marginal (both at the level of 4$\sigma$).

\begin{table*}
\caption{Fit results and profile parameters. The uncertainties are
given in parentheses and expressed in units of the least significant
figure. Notes: ($a$) the uncertainty
is half of that for the line FWHM, ($b$) the uncertainty is the spectral
resolution listed in Table~\protect\ref{obstab}, ($c$) the uncertainty
is the rms listed in Table~\protect\ref{spc_noise}, ($d$) the position
is the same as for the component above as they can not be distinguished in
the image, ($e$) the given upper limit is the spectral resolution, rather
than a FWHM estimate.}
\label{fit_results}
\begin{tabular}{@{}c@{\hskip 1mm}llrrrrlrrr}
\hline
     & \multicolumn{6}{c}{Gaussian components} & \cthead{Peak} &
\cthead{Peak} &\\
Spot & \cthead{LSR} & \cthead{$\alpha_{2000}$} & \cthead{$\delta_{2000}$}&
\cthead{Line} & \cthead{Flux} & \cthead{Size} & \cthead{LSR} &
\cthead{flux} &\cthead{$\int f(v)\;dv$}&\cthead{$T_b$} \\
     & \cthead{Velocity\makebox[0mm]{\hskip 2mm $^a$}} & \cthead{16$^h$58$^m$} & \cthead{$-$42\degr} &
\cthead{FWHM} & \cthead{density} & &
\cthead{velocity\makebox[0mm]{\hskip 2mm $^b$}} &
\cthead{density\makebox[0mm]{\hskip 2mm $^c$}} &
&\cthead{Limit}\\
     & \cthead{(\ks)} & \cthead{($^s$)}&\cthead{(arcmin~arcsec)} &
\cthead{(\ks)} & \cthead{(Jy)} & \cthead{(arcsec)} & \cthead{(\ks)} &
\cthead{(Jy)}&\cthead{(Jy \ks)} &\cthead{(K)} \\
\hline
A & \multicolumn{6}{c}{$5_{-1}-4_0$~E (84 GHz)} &
$-$32.82 & 14.6\hphantom{0} & 11.39\hphantom{0}~(7)& $4.5\times10^4$\\
  & $-$32.79\hphantom{0} & 16.585~(2) & 52~29.56\hphantom{1}~(2) &
0.72\hphantom{0}~(2) & 14.6~(3) & 0.40\hphantom{0}~(6) & \\
  & $-$31.83\hphantom{0} & 16.516~(8) & 52~28.42\hphantom{1}~(8) &
0.25\hphantom{0}~(9) & 0.8~(3) & 1.0\hphantom{00}~(3) & \\
  & \multicolumn{6}{c}{$8_0-7_1$~A$^+$ (95 GHz)} &
$-$32.80 & 38.80 &32.7\hphantom{00}~(1) & $9.5\times10^4$\\
  & $-$32.793 &  16.595~(1) & 52~29.74\hphantom{1}~(1) &
0.775~(6) & 38.6~(3) & 0.3\hphantom{00}~(2) & \\
  & $-$31.77\hphantom{0} & 16.527~(9) & 52~28.68\hphantom{1}~(8) &
0.28\hphantom{0}~(4) & 2.1~(3) & 0.9\hphantom{00}~(2) & \\

B & \multicolumn{6}{c}{$9_{-1}-8_{-2}$~E (9.9 GHz)} &
$-$31.56 & 9.5\hphantom{0} & 2.80\hphantom{0}~(4) & $5.3\times10^7$\\
  & $-$31.572 & 16.460~(2) & 52~25.73\hphantom{1}~(3) &
0.32\hphantom{0}~(1) & 8.1~(3) & 0.10\hphantom{0}~(9) & \\
  & $-$31.554\makebox[0mm]{\hskip 2mm $^d$} &  & &
\cthead{$<$0.029\makebox[0mm]{\hskip 2mm $^e$}} & 2.4~(3)& \\

  & \multicolumn{6}{c}{$2_2-2_1$~E (25~GHz)} &
$-$31.71 & 0.3\hphantom{0} & 0.032~(3) & $2.8\times10^4$\\
  & $-$31.69\hphantom{0} & 16.44\hphantom{0}~(2) &
52~26.4\hphantom{01}~(2) & 0.15\hphantom{0}~(8) & 0.2~(1) &
3\hphantom{.000}~(2) & \\

  & \multicolumn{6}{c}{$3_2-3_1$~E (25~GHz)} &
$-$31.61 & 5.9\hphantom{0} & 1.303~(8) & $2.6\times10^6$\\
  & $-$31.620 & 16.459~(2) & 52~25.90\hphantom{1}~(4) & 0.186~(7) &
5.2~(2) & 0.08\hphantom{0}~(3) & \\
  & $-$31.63\makebox[0mm]{\hskip 2mm $^d$}\hphantom{0} &  &  & 0.46~(16) &
0.6~(2) & \\

  & \multicolumn{6}{c}{$4_2-4_1$~E (25~GHz)} &
$-$31.71 & 8.7\hphantom{0} & 2.17\hphantom{0}~(6) & $3.9\times10^6$\\
  & $-$31.715 & 16.459~(2) & 52~25.90\hphantom{1}~(4) & 0.203~(7) &
8.3~(3) & 0.24\hphantom{0}~(6) & \\
  & $-$31.8\makebox[0mm]{\hskip 2mm $^d$}\hphantom{00} &  &  & 0.5\hphantom{00}~(3) &
0.6~(3) & \\

  & \multicolumn{6}{c}{$5_2-5_1$~E (25~GHz)} &
$-$31.69 & 6.7\hphantom{0} & 1.80\hphantom{0}~(2) & $3.0\times10^6$\\
 & $-$31.678 & 16.461~(8) & 52~25.70~(11) & 0.211~(5) &
6.4~(2) & 0.093~(8) & \\
 & $-$31.82\makebox[0mm]{\hskip 2mm $^d$}\hphantom{0} &  &  & 0.3\hphantom{00}~(1) &
0.5~(2) & \\
 & $-$31.24 & 16.581~(6) & 52~25.5\hphantom{01}~(1) &
0.5\hphantom{00}~(2) & 0.4~(2) & 0.5\hphantom{00}~(4) & \\

  & \multicolumn{6}{c}{$6_2-6_1$~E (25~GHz), May} &
$-$31.69 & 7.5\hphantom{0} & 2.047~(4) & $3.3\times10^6$\\
 & $-$31.701 & 16.464~(3) & 52~25.76\hphantom{1}~(5) & 0.211~(5) &
7.3~(2) & 0.17\hphantom{0}~(7) & \\
 & $-$31.89\makebox[0mm]{\hskip 2mm $^d$}\hphantom{0} &  &  & 0.18\hphantom{0}~(6) &
0.5~(2) & \\
 & $-$31.23 & 16.594~(1) & 52~25.32\hphantom{1}~(2) &
0.5\hphantom{00}~(1) & 0.6~(2) & 0.4\hphantom{00}~(1) & \\

  & \multicolumn{6}{c}{$6_2-6_1$~E (25~GHz), June} &
$-$31.67 & 7.2\hphantom{0} & 1.88\hphantom{0}~(6) & $6.4\times10^6$\\
  & $-$31.649 & 16.454~(1) & 52~25.70\hphantom{1}~(2) & 0.203~(9) &
6.8~(3) & 0.079~(7) & \\
  & $-$31.80\makebox[0mm]{\hskip 2mm $^d$}\hphantom{0} &  &  & 0.22\hphantom{0}~(9) &
0.8~(3) & \\
  & $-$31.2\hphantom{00} & 16.581~(1) & 52~25.25\hphantom{1}~(1) &
0.5\hphantom{00}~(3) & 0.5~(3) & 0.17\hphantom{0}~(5) & \\

  & \multicolumn{6}{c}{$7_2-7_1$~E (25~GHz)} &
$-$31.64 & 6.8\hphantom{0} & 1.77\hphantom{0}~(4) & $5.9\times10^6$\\
  & $-$31.660 & 16.462~(1) & 52~25.73\hphantom{1}~(1) &
0.24\hphantom{0}~(1) & 6.6~(3) & unresolved & \\
  & $-$31.3\hphantom{00} & 16.592~(2) & 52~25.26\hphantom{1}~(3) &
0.3\hphantom{00}~(2) & 0.4~(3) & 0.28\hphantom{0}~(5) & \\

  & \multicolumn{6}{c}{$8_2-8_1$~E (25~GHz)} &
$-$31.67 & 5.6\hphantom{0} & 1.48\hphantom{0}~(2) & $4.9\times10^6$\\
  & $-$31.644 & 16.461~(3) & 52~25.78\hphantom{1}~(5) &
0.24\hphantom{0}~(1) & 5.5~(3) & 0.060~(8) & \\

  & \multicolumn{6}{c}{$9_2-9_1$~E (25~GHz)} &
$-$31.63 & 3.9\hphantom{0} & 0.956~(4) & $3.3\times10^6$\\
  & $-$31.62\hphantom{0} & 16.458~(1) & 52~25.70\hphantom{1}~(1) &
0.24\hphantom{0}~(2) & 3.6~(3) & 0.06\hphantom{0}~(2) & \\

  & \multicolumn{6}{c}{$5_{-1}-4_0$~E (84 GHz)} &
$-$31.58 & 12.3\hphantom{0} & 5.35\hphantom{0}~(2) & $3.7\times10^4$\\
  & $-$31.85\hphantom{0} & 16.470~(8) & 52~26.29\hphantom{1}~(6) &
0.38\hphantom{0}~(3) &  2.5~(2) & 0.8\hphantom{00}~(4) & \\
  & $-$31.561 & 16.454~(2) & 52~25.72\hphantom{1}~(2) & 0.215~(5) &
9.3~(2) & 0.46\hphantom{0}~(8) & \\
  & $-$31.31\hphantom{0} & 16.470~(5) & 52~25.73\hphantom{1}~(4) &
0.59\hphantom{0}~(4) & 3.4~(2) & 0.68\hphantom{0}~(8) & \\

  & \multicolumn{6}{c}{$8_0-7_1$~A$^+$ (95 GHz)} &
$-$31.52 & 27.90 & 19.0\hphantom{00}~(1) & $6.8\times10^4$\\
  & $-$32.35\hphantom{0} & 16.469~(2) & 52~26.52\hphantom{1}~(2) &
0.59\hphantom{0}~(6) & 2.8~(2) & 0.4\hphantom{00}~(1) & \\
  & $-$31.808 & 16.467~(5) & 52~26.15\hphantom{1}~(4) & 0.415~(9) &
7.2~(2) &  0.7\hphantom{00}~(1) & \\
* & $-$31.516 & 16.463~(1) & 52~25.83\hphantom{1}~(1) & 0.329~(3) &
20.3~(2) & 0.30\hphantom{0}~(2) & \\
  & $-$31.235 & 16.473~(3) & 52~25.86\hphantom{1}~(3) & 0.594~(9) &
10.4~(2) & 0.60\hphantom{0}~(5) & \\
  & $-$30.56\hphantom{0} & 16.48\hphantom{0}~(1) &
52~26.1\hphantom{01}~(1) &
0.46\hphantom{0}~(8) & 1.0~(2) & 1.3\hphantom{00}~(5) & \\

  & \multicolumn{6}{c}{$11_{-1}-10_{-2}$~E (104 GHz)} &
$-$31.60 & 9.59 & 2.348~(3) & $2.0\times10^4$\\
  & $-$31.594 & 16.462~(2) & 52~25.64\hphantom{1}~(2) &
\cthead{$<$0.022\makebox[0mm]{\hskip 2mm $^e$}}& 2.2~(1) & 0.2\hphantom{00}~(1) & \\
  & $-$31.580\makebox[0mm]{\hskip 2mm $^d$} &  &  &
0.293~(5) & 7.5~(1) &  \\

%
\hline
\end{tabular}
\end{table*}

\begin{table*}
\contcaption{}
\begin{tabular}{@{}c@{\hskip 1mm}llrrrrlrrr}
\hline
     & \multicolumn{6}{c}{Gaussian components} & \cthead{Peak} &
\cthead{Peak} &\\
Spot & \cthead{LSR} & \cthead{$\alpha_{2000}$} & \cthead{$\delta_{2000}$}&
\cthead{Line} & \cthead{Flux} & \cthead{Size} & \cthead{LSR} &
\cthead{flux} &\cthead{$\int f(v)\;dv$}&\cthead{$T_b$} \\
     & \cthead{Velocity} & \cthead{16$^h$58$^m$} & \cthead{$-$42\degr} &
\cthead{FWHM} & \cthead{density} & & \cthead{velocity} & \cthead{density} &
&\cthead{Limit}\\
     & \cthead{(\ks)} & \cthead{($^s$)}&\cthead{(arcmin~arcsec)} &
\cthead{(\ks)} & \cthead{(Jy)} & \cthead{(arcsec)} & \cthead{(\ks)} &
\cthead{(Jy)}&\cthead{(Jy \ks)} &\cthead{(K)} \\
\hline

C & \multicolumn{6}{c}{$5_{-1}-4_0$~E (84 GHz)} &
$-$30.77 & 2.1\hphantom{0} & 0.7\hphantom{00}~(1) & $6.0\times10^3$\\
  & $-$31.14\hphantom{0} & 16.71\hphantom{0}~(1) &
52~20.7\hphantom{0}~(1) & 0.30\hphantom{0}~(5) &
1.2~(2) & 1.0\hphantom{0}~(1) & \\
  & $-$30.77\hphantom{0} & 16.672~(2) & 52~20.62~(2) &
0.28\hphantom{0}~(4) & 1.3~(2) & 0.4\hphantom{0}~(2) & \\
  & \multicolumn{6}{c}{$8_0-7_1$~A$^+$ (95 GHz)} &
$-$30.76 & 7.92 & 3.549~(3) &$1.9\times10^4$\\
  & $-$31.46\hphantom{0} & 16.75\hphantom{0}~(1) &
52~20.7\hphantom{0}~(1) & 0.32\hphantom{0}~(5) &
0.7~(1) & 2.0\hphantom{0}~(5) & \\
  & $-$31.167 & 16.707~(2) & 52~20.85~(1) & 0.270~(8) &
3.5~(1) & 0.36~(5) & \\
  & $-$30.747 & 16.686~(2) & 52~20.79~(1) & 0.276~(4) &
7.8~(1) & 0.36~(3) & \\

D & \multicolumn{6}{c}{\makebox[0mm]{\hskip -4cm (marginal detection)}$5_{-1}-4_0$~E (84 GHz)} &
$-$30.77 & 3.0\hphantom{0} & 0.8\hphantom{00}~(2) & \\
  & $-$30.8\hphantom{00} & & & 0.6\hphantom{00}~(2) & 1.5~(5) & \\
  & \multicolumn{6}{c}{$8_0-7_1$~A$^+$ (95 GHz)} &
$-$30.51 & 9.5\hphantom{0} & 7.54\hphantom{0}~(4) &$1.7\times10^4$\\
  & $-$30.92\hphantom{0} & 16.82\hphantom{0}~(1) &
51~59.63~(9) & 0.29\hphantom{0}~(3) & 5.3~(5) 
& 0.7\hphantom{0}~(6) & \\
  & $-$30.45\hphantom{0} & 16.850~(7) & 51~59.98~(5) &
0.62\hphantom{0}~(4) & 9.0~(5) & 0.7\hphantom{0}~(3) & \\

E & \multicolumn{6}{c}{\makebox[0mm]{\hskip -4cm (marginal detection)}$5_{-1}-4_0$~E (84 GHz)} &
$-$30.33 & 1.5\hphantom{0} & 0.4\hphantom{00}~(1) & \\
  & $-$32.18\hphantom{0} &  &  & 0.09\hphantom{0}~(5) & 0.5~(3) & \\
  & $-$30.336\hphantom{0} & &  & 0.05\hphantom{0}~(1) & 1.2~(3) & \\
  & $-$30.2\hphantom{00} & & & 0.9\hphantom{00}~(7) & 0.3~(3) & \\
  & \multicolumn{6}{c}{$8_0-7_1$~A$^+$ (95 GHz)} &
$-$30.41 & 7.1\hphantom{0} & 3.29\hphantom{0}~(9) & $1.6\times10^4$\\
  & $-$33.28\hphantom{0} & 17.53\hphantom{0}~(1) &
52~05.14~(9) & 0.16\hphantom{0}~(3) & 1.1~(2)
& 1.0\hphantom{0}~(5) & \\
  & $-$31.40\hphantom{0} & 17.72\hphantom{0}~(1) &
52~03.80~(8) & 0.12\hphantom{0}~(4) &
0.5~(2) & 2\hphantom{.00}~(2) & \\
  & $-$31.16\hphantom{0} & 17.48\hphantom{0}~(2) &
52~03.9\hphantom{0}~(2) & 0.19\hphantom{0}~(5) &
0.8~(2) & 1.4\hphantom{0}~(3) & \\
  & $-$30.403 & 17.549~(4) & 52~04.09~(3) & 0.210~(8) &
5.1~(2) & 0.62~(3) & \\
  & $-$30.185 & 17.551~(2) & 52~04.30~(1) &
0.32\hphantom{0}~(1) & 5.2~(2) & 0.38~(7) & \\

F & \multicolumn{6}{c}{$5_{-1}-4_0$~E (84 GHz)} &
$-$27.28 & 20.4\hphantom{0} & 9.3\hphantom{00}~(2) &$5.8\times10^4$\\
  & $-$27.87\hphantom{0} & 16.2\hphantom{00}~(1) &
52~00\hphantom{.00}~(1) & 0.18\hphantom{0}~(5) &
1.7~(4) & unresolved & \\
  & $-$27.436 & 16.163~(7) & 52~01.02~(6) &
0.37\hphantom{0}~(1) & 11.1~(4) & 0.8\hphantom{0}~(3) & \\
  & $-$27.311 & 16.187~(8) & 52~00.99~(6)  & 0.106~(5) &
9.8~(4) & 0.8\hphantom{0}~(2) & \\
  & $-$27.108 & 16.146~(5) & 52~00.73~(4) & 0.254~(9) &
12.8~(4) & 0.6\hphantom{0}~(3) & \\
  & \multicolumn{6}{c}{$8_0-7_1$~A$^+$ (95 GHz)} &
$-$27.14 & 36.6\hphantom{0} & 20.56\hphantom{0}~(7) &$8.7\times10^4$\\
  & $-$27.94\hphantom{0} & 16.190~(7) & 52~01.14~(6) &
0.10\hphantom{0}~(2) & 2.0~(3) & 0.9\hphantom{0}~(4) & \\
  & $-$27.75\hphantom{0} & 16.149~(4) & 52~01.11~(3) &
0.16\hphantom{0}~(2) & 3.5~(3) & 0.5\hphantom{0}~(2) & \\
  & $-$27.321 & 16.171~(3) & 52~01.23~(2) & 0.482~(5) &
30.3~(3) & 0.53~(5) & \\
  & $-$27.094 & 16.156~(1) & 52~01.13~(1) & 0.238~(5) &
16.9~(3) & 0.36~(4) & \\
\hline
\end{tabular}
\end{table*}

The profiles look similar at both frequencies for spot A. However, the
two spectral features of spot C show significantly different 95-GHz
peak flux densities, while at 84~GHz they are similar. The 84-GHz profile
of spot D looks somewhat similar to its 95-GHz counterpart, giving us
confidence that this marginal detection is real. The case for spot E
is more doubtful, a narrow spike near $-$30.33~\kss at 84~GHz may be
real given the presence of a feature with a similar appearence at
95~GHz. The 84-GHz spectrum also shows a hint of broad line emission
centred at $-$30.2~\kss, with a narrow line at $-$32.18~\kss which
does not have a 95-GHz counterpart.  We have formed similar spectra
for adjacent areas in the image to check that these 84-GHz features
are present towards the spots D and E only.  However, we can not
completely exclude the possibility that they are spurious because
spots D and E are located near the edge of the mosaic, and the source
contains emission at different locations with overlapping velocity
ranges.  In the image cube these features can not be distinguished
from the edge noise. Therefore, we have not drawn any conclusions from
the possible presence of 84-GHz emission towards spots D and E. The
95-GHz spectrum of spot E contains several components well spread in
velocity, although the image of the spot E is simple (see
Fig.~\ref{genmorph}).  This is not observed for any of the other spots
in this source.  The presence of multiple components for spot F is
more pronounced at 84~GHz. The 95-GHz profile of this spot looks like
a single asymmetric broad feature. This may indicate fine scale
structure, which differs at these two frequencies (i.e. there may be
one more 95-GHz component).

The remaining spot B, is exceptional in the sense that all observed
transitions have been detected towards it.  Corresponding spectra are
shown in Fig.~\ref{spotBsp}. As in Fig.~\ref{spotAC_Fsp}, the error
bars show the $3\sigma$ velocity uncertainty, with the exception of
the high J 25-GHz transitions where the uncertainty is less than the
spectral resolution, and therefore is not shown. The 95-GHz profile is
wider than the 84-GHz one and has an additional emission component. In
contrast, all 25-GHz profiles are narrow. The 9.9- and 104-GHz
profiles resemble each other and have an almost single spectral
channel spike on top of a broad symmetric line.

To provide quantitative characteristics of the observed spectral
profiles each of them has been fitted with a number of Gaussian
components. The results of these fits are summarized in
Table~\ref{fit_results}. As before, the uncertainties are given in
parentheses and expressed in units of the least significant
figure. The first column represents the name of the spot. The next six
columns are self-explanatory and contain the parameters of each
Gaussian component. The spectral and spatial domain were fitted
separately.  First, the peak radial velocity with respect to the local
standard of rest (LSR), the line widths (FWHM) and the peak flux
densities listed in the columns 2, 5 and 6 were determined directly
from the spectra shown in Fig.~\ref{spotAC_Fsp} and
\ref{spotBsp}. Then, a 2D Gaussian source was fitted to the image of a
single plane of the spectral cube, which corresponded to the peak
velocity of each Gaussian component. This second fit gave the position
and the deconvolved source size estimate shown in columns 3, 4 and
7. The latter is obtained under the assumption that the synthesized
beam is known exactly and the source is circular. In the size column
of Table~\ref{fit_results}, two components are described as
`unresolved' (the fitted size in these cases was slightly smaller than
the beam, owing to the effect of noise).  Marginal 84-GHz detections
towards the spots D and E are hidden in the edge noise in the
image. Therefore, no second stage fit has been done for them. However,
the position of these 84-GHz features, if they are real, is within
4\arcsec{ }(the box size used in forming the spectra from the image
cube) of the position of the appropriate 95-GHz spot. A combination of
a $\delta$-function and a Gaussian was used to fit the 9.9- and
104-GHz spectra, which contain a narrow spike. An upper limit equal to
the spectral resolution is given in the line width column of
Table~\ref{fit_results} for such narrow spikes. The flux density of
these spikes exceeds the rms noise 8 and 22 times for the 9.9 and
104~GHz profiles respectively (see Table~\ref{fit_results}).  This
fact leaves no doubts that the spikes are real.

All uncertainties listed in columns 3-7 are the formal uncertainties
of the Gaussian fit ($1\sigma$). The formal uncertainty of the peak velocity is
always half of that for the line width. However, due to blending of
different components seen in most cases, there are likely to be
systematic errors. Due to these errors, the individual
components are likely to appear closer to each other according to the
fit (given in Table~\ref{fit_results}) than they are in reality, both in the spatial
domain and velocity. Their line
widths and sizes are likely to be overestimated. In some cases, a
superposition of two components is completely dominated by the
brightest feature located at approximately the same velocity as the
weaker one. We were unable to produce a reliable image plane fit for a
weaker line in such situations.

The rest of the columns describe each line profile as a whole,
providing more robust values than the results of the Gaussian fit.
The columns are the peak LSR velocity and corresponding flux density,
the area under the line profile (the integrated flux density), and the
brightness temperature lower limit. The latter is calculated using the
peak (in the spectral domain) of the total brightness contained in the
same region of image, which was used to compute the spectra. This
approach is more appropriate for unresolved or barely resolved sources
and does not depend on (sometimes very crude) size estimates.  Such
high values of the brightness temperature (lower limits from
$6\times10^3$ to $5\times10^7$~K, see Table~\ref{fit_results}) are
generally considered as an indication that the detections have a maser
rather than thermal origin.  The uncertainty of the peak velocity is
equal to the spectral resolution, which is different in each
particular case (see Table~\ref{obstab}). A systematic offset of the
velocity scale for a transition can be caused by rest frequency
uncertainties, which are also listed in Table~\ref{obstab}. This also
applies to the peak velocity of the Gaussian components. The
uncertainty of the integrated flux density was obtained from either
the noise level in the spectrum (see below) or the offset from zero of
the mean flux density in the line-free part of the spectrum; we have
quoted the larger of these two uncertainties in each case.

To quantify the noise and provide useful upper limits for transitions
which have not been detected we have calculated the rms noise in the spectra
constructed for each spot. The spectra for non-detections were
extracted from the cube in essentially the same way as the spectra
with a detected line emission shown in Fig.~\ref{spotAC_Fsp} and
\ref{spotBsp}.  For the spectra with a detected emission, rms
calculations were confined to the emission free part of the
spectrum. The $1\sigma$ noise levels are listed in
Table~\ref{spc_noise}. They vary from spot to spot for the high
frequency transitions due to the small primary beam size (comparable
with the spread of the spots in the source) and the lack of an
adequate mosaicing in the North-South direction. The noise levels list
in Table~\ref{spc_noise} are the uncertainties of the peak flux
densities in Table~\ref{fit_results} (8th column).

\subsection{Notes on temporal variability}
\label{tempvar_section}

Methanol masers are known to be variable
\citep*[e.g.][]{cas95a,goe04}, although the subject is not very well
studied for class~I masers. One of the goals of this work is to
provide data for further development of maser models. The models are
usually checked by comparing the flux density ratios of different
transitions. However, these ratios can be affected not only by the
studied pumping mechanism, but also by any temporal variability, if
observations are not simultaneous. Temporal variability is typically
not included in modelling and requires a long time series of data to
be well understood. To obtain some estimate of the scale of temporal
variability in the source we repeated the observations of some
transitions in different observing runs, in two cases (see
Table~\ref{obstab}) without a proper imaging (a single short
scan). Provided the position of the source is known from another
observing run, such non-imaging data still allows us to estimate the
peak flux density and, therefore, to assess the variability.  We have
performed calibration similar to that used for the main dataset and
obtained the following estimates for the peak flux density by fitting
a point source at the known position to the visibility data:
6.7$\pm$0.2~Jy for the 25-GHz J=6 data taken in the August session,
and 9.0$\pm$0.2~Jy for the 9.9-GHz data taken in May. The 25-GHz J=6
transition was observed in all three sessions, and the above value of
the flux density is in agreement with that in Table~\ref{fit_results}
within $3\sigma$ uncertainty. The 9.9-GHz non-imaging data are in
agreement with the corresponding imaging experiment as well. The
latter is also important for our conclusion that polarization effects
do not significantly affect the results of our observations
(section~\ref{obs_section}).  This is because, in contrast to the June
observations, the 9.9-GHz non-imaging observations carried out in May
used a dual polarization mode and provided effectively a Stokes-I
measurement.  Taking into account the systematic errors in
establishing the absolute flux density scale, which are estimated to
be around 10\% and 3\% at 25 and 9.9~GHz, respectively, we can
conclude any temporal variability within the period of our
observations is less than the calibration errors. Therefore, we can
consider that we have inputs for the modelling which are free from any
significant variability bias.

\section{Discussion}
\subsection{Association with outflow}
\begin{table}
\caption{The noise level (1-$\sigma$-rms) in the spectra.}
\label{spc_noise}
\begin{tabular}{@{}l@{\hskip 2mm}rrrrrr}
\hline
\cthead{Transition}&\multicolumn{6}{c}{Maser spot}\\
\cthead{and}&\cthead{A} & \cthead{B} & \cthead{C} & \cthead{D} &
\cthead{E} & \cthead{F} \\
\cthead{frequency}&\multicolumn{6}{c}{1-$\sigma$ rms (Jy)}\\
\hline
$9_{-1}-8_{-2}$~E\hfill(9.9 GHz) & \multicolumn{6}{c}{0.3 for all spots}\\
$\mathrm{J}_2-\mathrm{J}_1$~E, May\hfill(25~GHz)  &\multicolumn{6}{c}{0.1 for all spots}\\
$\mathrm{J}_2-\mathrm{J}_1$~E, June\hfill(25~GHz) &\multicolumn{6}{c}{0.2 for all spots}\\
$5_{-1}-4_0$~E\hfill(84 GHz) & 0.2 & 0.2 & 0.2 & 0.5 & 0.3 & 0.5\\
$8_0-7_1$~A$^+$\hfill(95 GHz)& 0.09 & 0.09 & 0.07 & 0.3 & 0.2 & 0.4\\
$11_{-1}-10_{-2}$~E~~\hfill(104 GHz) & 0.1 & 0.07 & 0.08 & 0.4 & 0.3 & 0.5\\
\hline
\end{tabular}
\end{table}

All the detected maser spots are located within a dense molecular core
\citep[deconvolved FWHM 27\arcsec,][]{gar03} and have velocities close
to that of the ambient molecular material in the core ($-$30.0~\ks).
A remarkable correlation between the southern maser spots (labelled A
to C) and the shocked gas mentioned in section \ref{genmorph_section}
(see also Fig.~\ref{genmorph}) points to an association of the class~I
masers with the interface region of this outflow.  This result is
somewhat similar to that obtained for the northern star-forming region
DR~21 by \citet{pla90} more than a decade ago. In addition, \citet{kur04}
revealed the same association for a few other northern sources. However
a more favourable geometry and probably lesser complexity in the IRAS
16457-4247 star-forming region allows us to study this association in
detail.

In contrast to the southern spots, no associated H$_2$ emission has
been detected in the vicinity of the northern spots E and F (see
Fig.~\ref{genmorph}). However, spot D resides close to the northern
continuum lobe, which is associated with faint H$_2$ emission. This
morphology leaves no doubts that spot D is very similar to the
southern maser spots, because the continuum source represents a
working surface of the jet \citep{bro03}. Moreover, it is evident from
the large-scale image of the H$_2$ emission (Fig.~\ref{genmorph}) that
an outflow lobe extending northwards is present.  Therefore it is
likely that all maser spots in this source are associated somehow with
the molecular outflow. Two factors or their combination can explain
the non-detection of the H$_2$ emission in the vicinity of the maser
spots E and F. First, the 2.12~$\mu$m H$_2$ emission can be affected
by extinction, which may be higher in the northern part of the
region. Such variations of extinction have been observed in other
sources \citep*[e.g.,][]{eis00,nan02}. In addition, \citet{gar06} have
recently discovered a bipolar molecular outflow towards IRAS
16457-4247.  This outflow is inclined to the line of sight at an angle
of 40\degr{ }with the receding flow located in the northern part of
the region. A higher extinction to the north from the central object
has a natural explanation in this geometry.

Alternatively, or in addition to this, the conditions in the interface
region may be less energetic in the northern part of the source, if
for example most of the dense material blocking the path of the
outflow has already been swept away. The cooling timescale of the
gas radiating shocked H$_2$ emission is very short and may be of the
order of several years \citep[e.g.,][]{dav95,smi90}. This estimate is
in agreement with measurements of variability performed by
\citet{mic98} for the HH~46/47 outflow system. Because the IRAS
16457-4247 outflow extends well beyond the region containing maser
spots (see Fig.~\ref{genmorph}), the shocked gas in the vicinity of
northern maser spots generated when the jet carved out a cavity in the
dense molecular core, has already had enough time to radiate the
energy and cool down if the interaction became less energetic for some
reason. It is reasonable to expect that the shocks associated with
internal working surfaces (where a fast moving gas impacts a slow
moving gas and drives shocks into the medium) are less energetic than
those associated with the terminal working surfaces (where the flow
impacts the cloud directly). To the south from the central source, the
jet may have impacted the walls of the cavity generating a young
curved shock. However, as it was mentioned above, there are a number
of other factors that can account for the shape of this H$_2$
feature. It must be noted that this scenario is speculative as there
is no direct observational evidence at this stage that suggests a
weaker interaction of the northern flow with the cloud.

The main role of shocks for methanol maser formation is thought to be
that they increase the methanol abundance in the gas phase. The
production of methanol by purely gas-phase chemical reactions is
inefficient, but the passage of weak shocks is able to release it from
the grain mantles \citep[e.g.,][]{har95}.  Shocks generated due to the
interaction between outflows and the ambient material are known to
enhance the methanol abundance \citep[e.g.,][]{gib98,gar02}. In
addition, outflows compress the medium, which makes collisions more
frequent and the pumping mechanism more efficient. The shocks have to
be relatively gentle as methanol can survive sputtering or desorption
of grain mantles only at shock velocities not greatly exceeding
10~\kss\citep{gar02}.  Therefore, class~I methanol masers are likely to
reside in the wakes associated with the bow-shocks or internal working
surfaces rather than near the apex of a terminal bow-shock.  The fact
that all maser spots are only found close to the jet (Fig.~\ref{genmorph})
supports this idea. 

It is worth noting that spot B, which is active in all observed
transitions, is located near the edge of the brightest H$_2$ knot.  It
is the only spot in this source where maser emission at 9.9, 25 and
104~GHz has been detected. According to models calculated by
\citet{sob05} these masers require more energetic conditions (higher
densities and temperatures) than the common masers at 84 or 95~GHz.
Our results confirm this model, although with the caveat that the observed
H$_{2}$ emission may be affected by variable extinction. Spectroscopy
of a number of H$_2$ transitions, as well as observations of other
shock tracers less prone to extinction (e.g. SiO and thermal methanol
transitions) are required before firm conclusions can be drawn.
The 9.9- and 104-GHz profiles contain a narrow spike (Fig.~\ref{spotBsp})
which has a width comparable to the spectral resolution ($<$0.03~\ks,
see Table~\ref{obstab}), and is the narrowest spectral feature ever claimed
according to our knowledge. Such a narrow feature implies that the maser
is unsaturated and requires turbulent motions
to be suppressed (e.g. by the magnetic field) in the gas involved in
the maser action. 

The clear association of class~I methanol masers with outflows in this
and other sources suggests that searches for class I methanol masers
should be extended to sites of intermediate- and low-mass star
formation, where outflows are common \citep{rei01}.  Indeed, at such
sites a few class I masers have recently been found \citep{kal06}, but
a lack of accurate positions for these masers means their association
with outflows cannot be verified at the present.

\subsection{Masers at 84 and 95~GHz}
\begin{figure}
\includegraphics[width=\linewidth]{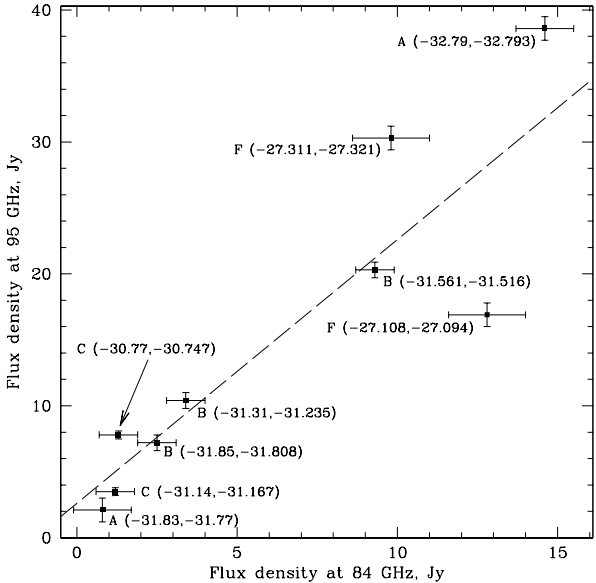}
\caption{The correlation between flux densities at 84 and 95~GHz. Each
data point is labelled with the spot name and velocities in \kss
(two numbers in brackets) corresponding to the 84- and 95-GHz components
taken from Table~\ref{fit_results}. The error bars correspond to the 3$\sigma$
formal uncertainty of the Gaussian fit. The dashed line with the slope
2.1$\pm$0.3 represents the least square fit discussed in the text.}
\label{spots84_95}
\end{figure}

As mentioned above, the maser profiles at 84 and 95~GHz look similar
in general (Fig.~\ref{spotAC_Fsp} and \ref{spotBsp}). This suggests a
common range of physical conditions required for these masers to
form. \citet{val00} found a similar correspondence between the 95-GHz
masers and the $7_0-6_1$~A$^+$ masers at 44~GHz (this frequency is
currently outside the band of the ATCA receivers and has not been
observed in this project) using a large sample of southern maser
sources, including IRAS~16547-4247. The flux densities of these
two transitions were found to correlate,
despite a significant scatter of the points, which is most likely
caused by the lack of accurate positions for the majority of known
class~I methanol masers.  However, the 44- and 95-GHz masers belong to
the same $(\mbox{J}+1)_0-\mbox{J}_1$~A$^+$ series of transitions
(J=6 and 7 respectively) and are expected to correlate. In contrast,
the 84 and 95-GHz maser transitions belong to a different series of
transitions, and even to different species of methanol (E- and
A-methanol respectively, see Table~\ref{obstab}), which act as two
different molecules. Therefore, a correlation between the flux densities
corresponding to these two transitions is not self-evident and can be
a constraining factor for models of maser pumping.

Fig.~\ref{spots84_95} shows the peak flux densities ($F$) at 84 and 95~GHz
for matching pairs of the Gaussian components listed in
Table~\ref{fit_results}.  Each point is labelled with the spot name
and the peak velocities of the corresponding 84- and 95-GHz Gaussian
components. The error bars show $3\sigma$ formal uncertainties of the
fit and do not take into account possible systematic errors due to the
absolute flux density scale and the structure of each spectral
profile. It was concluded from the general appearance of the profiles
that the spot F is likely to have an additional component at
95~GHz. Therefore, the systematic errors are particularly acute for
this spot. Despite this caveat, Fig.~\ref{spots84_95} reveals the
correlation between flux densities quite well. A linear least square
fit gives the following dependence
\begin{equation}
\label{corr95_84}
F(\mbox{95~GHz})=(2.1\pm0.3)\times F(\mbox{84~GHz})+(2.6\pm2.5)\mbox{,}
\end{equation}
which is shown in Fig.~\ref{spots84_95} by the dashed line. The
additive term is not statistically significant. The absolute flux
density scale calibration affects all spots at the same frequency in
the same way.  Therefore, the slope has an additional uncertainty of
30\% (about 0.6).  \citet{kal01} obtained a similar result to
(\ref{corr95_84}) by comparing single dish surveys of a large number
of sources at these two frequencies.  However, to study such a
dependence reliably, an interferometer is required, due to the spread
of maser spots. It is worth mentioning that interferometric
observations of the DR21 star-forming complex carried out by
\citet{bat88} and \citet{pla90}, found the ratio of the 95- to 84-GHz
flux density to be approximately 5.5, somewhat higher than observed in
IRAS 16547-4247.

\subsection{Rest frequencies}

High spectral resolution observations can be used to improve the
accuracy to which the transition rest frequencies are known. However,
in many cases, the laboratory measurement is preferred regardless of
its accuracy because the assumption of equal velocities for all
transitions may be misleading (e.g., the discussion about the 95- and
25-GHz masers in the next section). According to
Fig.~\ref{spotBsp} the 9.9- and 104-GHz transitions have a single
channel spike on top of the broad line. These two transitions belong
to the same $(\mbox{J+1})_{-1}-\mbox{J}_{-2}$~E transition series (J=8
and 10) and are expected to show a similar behavior. Therefore,
assuming that the spike is at the same velocity, the rest frequency of
the $9_{-1}-8_{-2}$~E transition is 9936.201$\pm$0.001~MHz. The
uncertainty is determined by the spectral resolution and the rest
frequency uncertainties of the 9.9- and 104-GHz transitions.
The rest frequency was adjusted with respect to Table~\ref{obstab}
by 1~kHz or 0.03~\ks. It is not practical to adjust
the rest frequency for other transitions.

\subsection{Maser models}
\begin{table*}
\caption{Model results for spot B. $R$ is the ratio of the integrated flux
density to that of the J=4 transition at 25~GHz, $\tau$ is the optical
depth and $T_b$ is the brightness temperature. The uncertainties of the observed
$\log(R)$ are given in the parentheses and expressed in units of the least
significant figure. According to
Table~\ref{fit_results} the brightness temperature of the J=4 25-GHz transition
exceeds $3.9\times10^6$~K.}
\label{model_res}
%
\begin{tabular}{@{}l@{\hskip 4mm}rrcrrcrr}
\hline
\cthead{Transition}&\cthead{$\nu$}&\cthead{Observed}&\multicolumn{3}{c}{Model 1}&\multicolumn{3}{c}{Model 2}\\
&(GHz)&\cthead{$\log(R)$}&\cthead{$\log(T_b, \mbox{K})$} & \cthead{$\tau$} & \cthead{$\log(R)$} &
\cthead{$\log(T_b, \mbox{K})$} & \cthead{$\tau$} & \cthead{$\log(R)$} \\
\hline
$\hphantom{1}9_{-1}-\hphantom{1}8_{-2}$~E&9.9& 0.11~(7) &8.92 & $-$17.9 & $-$0.1 & 8.81 & $-$19.1 & 0.5\\
$\hphantom{1}2_{2\hphantom{-}}-\hphantom{1}2_{1\hphantom{-}}$~E&25& $-$1.8\hphantom{0}~(1) & 6.34 & $-$10.1 & $-$1.9 & 5.29 & $-$9.9 & $-$2.2\\
$\hphantom{1}3_{2\hphantom{-}}-\hphantom{1}3_{1\hphantom{-}}$~E&25& $-$0.22~(4) & 7.98 & $-$13.7 & $-$0.3 & 7.36 & $-$14.6 & $-$0.2\\
$\hphantom{1}4_{2\hphantom{-}}-\hphantom{1}4_{1\hphantom{-}}$~E&25& 0\hphantom{.00~(0)} & 8.27 & $-$14.2 & 0\hphantom{.0} & 7.56 & $-$15.0 & 0\hphantom{.0}\\
$\hphantom{1}5_{2\hphantom{-}}-\hphantom{1}5_{1\hphantom{-}}$~E&25& $-$0.08~(8) & 8.34 & $-$14.3 & 0.1 & 7.55 & $-$14.9 & 0\hphantom{.0} \\
$\hphantom{1}6_{2\hphantom{-}}-\hphantom{1}6_{1\hphantom{-}}$~E\makebox[0mm]{\hskip 9mm (May)} &25& $-$0.03~(7) & 8.43 & $-$14.5 & 0.2 & 7.36 & $-$14.5 & $-$0.2\\
$\hphantom{1}6_{2\hphantom{-}}-\hphantom{1}6_{1\hphantom{-}}$~E\makebox[0mm]{\hskip 9mm (June)}&25& $-$0.1\hphantom{0}~(1) & \\
$\hphantom{1}7_{2\hphantom{-}}-\hphantom{1}7_{1\hphantom{-}}$~E&25& $-$0.09~(9) & 8.47 & $-$14.7 & 0.2 & 7.00 & $-$13.7 & $-$0.5\\
$\hphantom{1}8_{2\hphantom{-}}-\hphantom{1}8_{1\hphantom{-}}$~E&25& $-$0.17~(8) & 8.45 & $-$14.7 & 0.2 & 6.27 & $-$12.2 & $-$1.3\\
$\hphantom{1}9_{2\hphantom{-}}-\hphantom{1}9_{1\hphantom{-}}$~E&25& $-$0.36~(7) & 8.45 & $-$14.8 & 0.2 & 5.36 & $-$10.2 & $-$2.2\\
$\hphantom{1}5_{-1}-\hphantom{1}4_{0\hphantom{-}}$~E&84& 0.4\hphantom{0}~(1) & 2.38 & 20.9 & $-$4.8 & 6.95 & $-$12.3 & 0.5\\
$\hphantom{1}8_{0\hphantom{-}}-\hphantom{1}7_{1\hphantom{-}}$~A$^+$&95& 0.9\hphantom{0}~(1) & 1.95& 71.9 & $-$5.1 & 6.83 & $-$11.9 & 0.4\\
$11_{-1}-10_{-2}$~E&104 & 0.0\hphantom{0}~(1) & 2.93&1.2 & $-$4.1 & 5.58 & $-$10.7 & $-$0.7\\
\hline
\end{tabular}
\end{table*}

Theoretical studies of class~I methanol masers were impeded for a long
time by the absence of an adequate model of collisions for the
methanol molecule. Recently calculated rate coefficients
\citep*[e.g.,][]{pot04} have encouraged pumping studies of these
masers.  \citet{sob05} performed a qualitative analysis and revealed
that the 25-GHz masers are bright in models with specific
column density (equivalent to the column density divided by the
line-width) greater than 10$^{12}$~cm$^{-3}$~s and require relatively
high kinetic 
temperatures (75$-$100~K) and densities (10$^5-10^7$~cm$^{-3}$).
Similar conditions are required for the 9.9- and 104-GHz masers,
although they prefer a greater beaming (elongated geometry) and either
lower densities or higher temperatures (more than 100~K). Spot B is an
excellent test site to develop a class~I pumping model because of the
large number of transitions detected. However, a detailed
investigation of the vast parameter space of the maser models is
beyond the scope of this paper and will be reported elsewhere.  At
present we are unable to reproduce the relative brightnesses of all
transitions detected in spot B in a single model similar to that of
\citet{sob05}. We suspect that gradients in the physical parameters
may be influential and leading to the co-existence of the different
pumping regimes on spatial scales unresolved by ATCA. It is expected
that strong gradients may accompany shocks, although their effects on
the maser pumping are yet to be studied. According to
Table~\ref{fit_results} the peak components at 25 and 95~GHz are
displaced by not more than 0\farcs1. Most likely this displacement is
caused by a systematic error due to the structure of the 95-GHz maser
because no such displacement is observed for the 84-GHz maser. Taking
this uncertainty as an upper limit and assuming a distance to the
source of 2.9~kpc, the parameters must change on a scale around 300 AU
or less.

It is worth noting, however, that firm conclusions about any gradients
can only be made after an extensive parameter search. In this paper we
have just illustrated the situation using two representative models
selected from the range of models calculated by
\citet{sob05}. Outlines of the modelling procedure applicable to
class~I masers can be found in \citet{vor05b} and the code is
described by \citet{sut04} and references therein. Both models use a
kinetic temperature (T$_{\mathrm{gas}}$) of 75~K, an intrinsic
linewidth (to simulate blending of spectral lines) of 0.5~\ks, and a
methanol abundance relative to hydrogen of $10^{-5}$. The parameters
that differ between
these two models are the hydrogen number density ($n_H$), $10^7$ and
$10^6$~cm$^{-3}$, the specific column density (equivalent to the column
density divided by the line-width), $10^{12}$ and
$10^{10.5}$~cm$^{-3}$~s, and the beaming
factor~($\varepsilon^{-1}$), 20 and 30, for the first and the second
models respectively. The results of the simulations are shown in
Table~\ref{model_res}. Most of the columns are self explanatory, the
columns labelled $\log R$ give a decimal logarithm of the ratio of
the integrated flux density
of the transition to that of the J=4 25-GHz transition (as observed
and for each of the models). The uncertainties of the observed
$\log(R)$ are given in the parentheses and expressed in units of the least
significant figure. These are cumulative uncertainties, which, in general,
account for both the absolute flux density scale uncertainty and that of the
integrated flux density given in Table~\ref{fit_results}. The J=2 and 3
transitions were observed simultaneously with the reference (J=4) transition
in the same band (see section~\ref{obs_section}).
Therefore, the ratio $R$ is not affected by
the absolute flux density scale calibration for these two transitions.
Other columns contain the brightness
temperatures ($\log T_b$) and optical depths ($\tau$) predicted by the
two models.

\begin{figure}
\includegraphics[width=\linewidth]{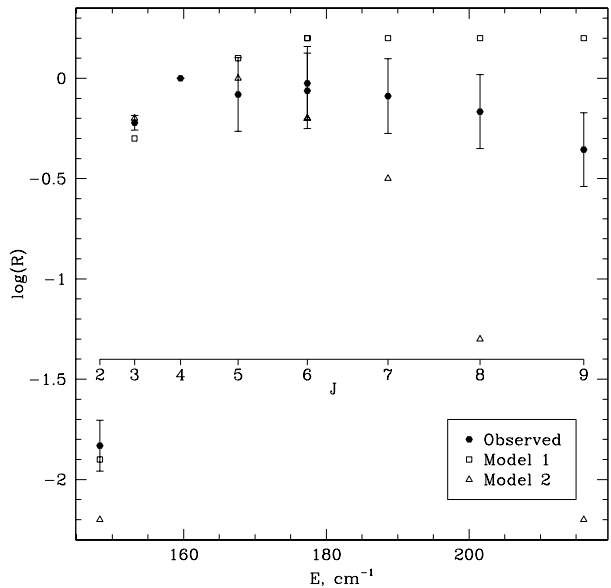}
\caption{The dependence of the observed and predicted relative integrated
flux densities ($R$) versus the exitation energy for the 25-GHz masers
detected in spot B. Open squares and triangles correspond to the first
and the second models (see Table~\protect\ref{model_res}), respectively.
Observed values are represented by the filled circles.
The J=4 transition is taken as a reference (see text). Therefore, the
observed point and the model points
coincide exactly for this transition. The error bars for observed values
represent a cumulative 3$\sigma$ uncertainty, which takes into account
the uncertainty of the absolute flux density scale for J$>$4
transitions (J=2,3 and 4 transitions were observed simultaneously in
the same band; see section~\ref{obs_section}).
There are two independent observations of the J=6 transition.
}
\label{rotd}
\end{figure}

Figure~\ref{rotd} illustrates the dependence of $\log R$ on the
excitation energy for the 25-GHz series (data from
Table~\ref{model_res}).  Open squares and triangles correspond to the
first and the second models, respectively. Observed values are shown
by the filled circles with error bars corresponding to the 3$\sigma$
uncertainty. The observed dependence represents essentially
a rotational diagram (because the product of the spontaneous decay
rate by the statistical weight does not vary much throughout the
series), which is a well known method to analyze the level populations
of the molecule \citep[e.g.][]{men86}. The plot of observed values
shows a significant curvature, although a linear fit to the highest J
transitions is possible. Such a fit, which takes into account all
transitions with J$>3$, has a rotational temperature of 80$\pm$10~K.
It should be remembered that this temperature is unlikely to be related
to any physical or even excitation temperature. The latter is
estimated correctly if the fitted 25-GHz transitions are all optically
thin or have the same optical depth.

It is evident from Table~\ref{model_res} that despite the fact that
the first model describes the ratios for the 9.9 and the 25-GHz
transitions quite well, it fails to produce the masers at 84, 95 and
104~GHz.  Similarly, the second model is adequate for 9.9-, 84-, 95- and
104-GHz masers, but fails to describe the behavior of the high J
25-GHz transitions (Fig.~\ref{rotd}). The brightness temperature of
the reference $4_2-4_1$~E transition is 1.9$\times10^8$
and $3.6\times10^7$~K, for the first and the second models
respectively (see Table~\ref{model_res}). Both these values are in
agreement with the observed lower limit ($3.9\times10^6$~K,
Table~\ref{fit_results}).

\subsection{Evolutionary stages and maser activity}
\label{evolution_section}

H$_2$O, OH and both classes of methanol masers often, but not always,
are present in the same region of high-mass star formation. The question
whether these masers trace an evolutionary sequence (with a partial
overlap in time) has been investigated in various studies
\citep*[e.g.,][]{for89,cod00,szy05,ell05,ell06}. From their analysis of
a large sample of star-forming regions \citet{for89} concluded that
the compact, isolated H$_2$O masers mark the earliest evolutionary
stages, and OH masers appear afterwards. The class~II methanol masers
at 6.7~GHz seem to largely overlap with the H$_2$O masers and precede
OH (e.g., Szymczak et al. \citeyear{szy05}). The majority of class~II
masers are not associated with detectable H{\sc ii} regions
\citep{phi98,wal98}, but show an association with millimetre and
submillimetre \citep{pes02,wal03} sources which leaves no doubt that
they trace a very early stage of the (proto)stellar evolution.  However,
the question as to where the class~I methanol masers fit into this
picture is still poorly understood. The main obstacle is the lack of
untargeted surveys of class~I masers, which have not been undertaken
mainly because there is no widespread, bright, low frequency maser
transition similar to the 6.7~GHz class~II transition, or mainline
OH. Most class~I masers, except for a few famous sources like OMC-1, have
been found towards known class~II methanol or OH maser sites, which
are often located within a single dish beam
\citep[e.g.,][]{sly94,ell05}. Therefore, all these samples are biased.
\citet{ell06} investigated this question using the infrared properties
of a subsample of methanol masers associated with GLIMPSE (Galactic
Legacy Infrared Mid-Plane Survey Extraordinaire) catalogue
point sources.  Despite the small number statistics, which makes
any firm conclusions impossible, it was found that the sources with a
class~I maser seem to have redder GLIMPSE colours than those without
it. Based on these results, \citet{ell06} concluded that the class~I
methanol masers may signpost an earlier stage of high-mass star
formation than the class~II masers.

\citet{for89} detected both mainline OH and H$_2$O masers in
IRAS 16547-4247 at a position close to the central continuum source in
Fig.~\ref{genmorph}. Taking into account the information given in the
previous paragraph, one may expect the source to be relatively old and
most likely to be active in class~II maser transitions. It is generally
assumed that each high-mass star-forming region has a detectable 6.7~GHz
maser sometime during the early stages of its evolution regardless of
the maser variability and the source geometry \citep[e.g.,][]{vdw05}.
However, no 6.7-GHz
emission has been detected in this source \citep[upper limit of
0.3~Jy,][]{cas95b,wal98}, in contrast to the detection of the large
number of class~I masers reported in this paper. This fact encourages
us to revisit the question of where class~I methanol masers fit into
an evolutionary sequence.

\begin{figure}
\includegraphics[width=\linewidth]{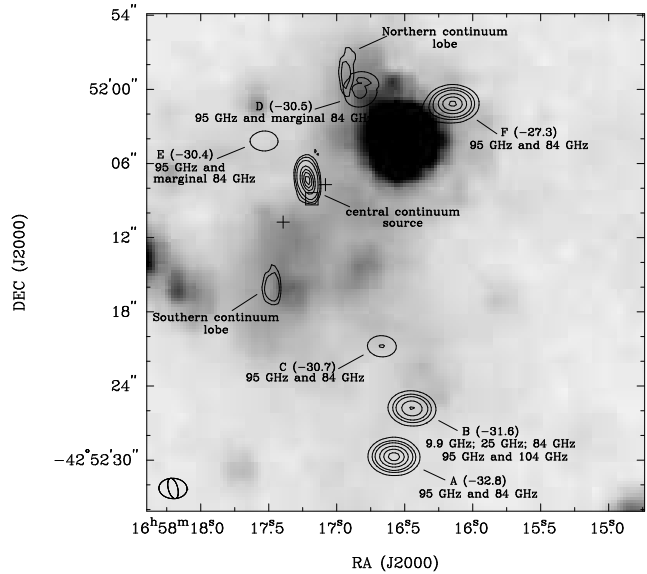}
\caption{The same as the inset in Fig.~\ref{genmorph}, but with the
GLIMPSE 5.8-$\mu$m emission shown by greyscale.}
\label{glimpse_mas}
\end{figure}

The GLIMPSE survey \citep{ben03} provides mid-infrared images of the
Galactic Plane with unprecedented sensitivity.
Figure~\ref{glimpse_mas} shows the locations of the radio continuum,
methanol, OH and water masers superimposed on the GLIMPSE 5.8-$\mu$m
emission (greyscale) on the same scale as Fig.~\ref{genmorph}.  The
most striking feature of this image is that there are no MIR sources
associated with either the jet source, nor the class~I masers.
Diffuse emission, probably arising from the Polycyclic Aromatic
Hydrocarbons (PAH) band at 6.2~$\mu$m appears to be associated with
the outflow region and traces a similar area to the shock-excited
H$_2$.  Although, unlike the H$_2$, the PAH emission is stronger
closer to the central continuum source and very faint in the vicinity
of spot~B.  The GLIMPSE 3.6-$\mu$m image shows similar diffuse
emission, also likely to be associated with PAH emission, in this case
from the 3.3-$\mu$m band.  Fig.~\ref{glimpse_mas} also shows a strong,
extended MIR source lying between spots D and F.  We have also
examined the {\em MSX} images of this region which show that although
the 12-$\mu$m emission (C-band) is centred on the strong MIR source
seen in Fig.~\ref{glimpse_mas}, the 21-$\mu$m emission (E-band) is
offset and centred on the central continuum source.  That the source
driving the jet/outflow/masers is only visible at wavelengths longer
than 12~$\mu$m suggests that this source is at a very early
evolutionary phase. \citet{gar03} argued that the observed
cm-wavelength continuum emission towards IRAS~16547-4247 is too weak
for such a luminous object ($6.2\times10^4$ L$_{\odot}$).  Supported
by the characteristics of the spectral line profiles, \citet{gar03}
explained this flux discrepancy by the ongoing intense accretion,
which forbids the development of an ultra-compact H{\sc ii} region.
Intense accretion is further evidence for an early evolutionary phase.
\citet{ell06} found the vast majority of class~II methanol maser
sources to be associated with MIR emission visible in the GLIMPSE
8.0-$\mu$m observations.  That such emission is absent in this source
argues that it is likely to be at an earlier evolutionary phase than
that which produces class~II methanol masers; however, this is in
direct contrast with the presence of OH masers which would generally
be taken to support the opposite interpretation.

In theory, a moderately dense gas (dense enough to switch the
collisional pumping on, but not dense enough to produce
thermalization), rich in methanol and separated from strong infrared
sources is able to produce most class~I methanol masers
\citep{cra92,vor99,vor05b}.  The fact that the class~I masers do not
require infrared radiation for pumping suggests that they might exist
at earlier evolutionary stages than the class~II masers, which need a
nearby infrared source for pumping.  However, a mechanism able to
increase the methanol abundance above that seen in cold cores must
already be active or have been active in the past to produce a
detectable maser. As was discussed above, a release of methanol from
the grain mantles by weak shocks is a good example of such a
mechanism. One cannot exclude the possibility that class I masers
survive even when the shock, which had increased the methanol
abundance in the past, has already dissipated.  Rare class~I masers at
9.9 and 104~GHz require more energetic conditions \citep{sob05}.  They
are most likely (or even exclusively) located near the young shocks.
Spot B in Fig.~\ref{genmorph} is likely to be a good example of such a
situation.  If the idea that the common class~I masers can outlast the
shock is correct, these masers become hard to destroy, once they
appear.  Situated at a considerable distance from the central object
they probably fade out slowly as the methanol abundance decreases
\citep[e.g. due to freezeout or gas-phase chemical
processing,][]{ave96}, unless the chemical composition of the medium
is altered once again by new energetic outflow events or by a slight
change in the outflow direction. A new epoch of interaction between
the outflow and the medium can either destroy the molecules engaged in
the maser emission or produce new masers. Assuming typical outflow
velocities of 10$^2-10^3$~\kss\citep[e.g.,][]{rei01},
it would take at least 10$^4$-10$^5$
years for this disturbance to reach the maser sites for the given
distance to the source (2.9~kpc) and the angular separations of masers
from the central object (see Table~\ref{fit_results}).  A timescale
for methanol destruction by ion-molecule reactions has approximately
the same order of magnitude \citep[e.g.,][]{san93,gar02}.

The class~II masers are usually located very close to the YSO
and, therefore, can be more easily destroyed during the course of its
evolution.  Given the detection of OH masers, which suggests an
evolved stage, it is, therefore, possible that the 6.7-GHz maser has
already been quenched in the studied source. We cannot presently
resolve the contradiction between the infrared data and the presence
of OH masers and confidently determine whether the source is too young
or too old to have a 6.7-GHz maser. However, these speculations show
that the evolutionary stage where the class~I masers are active, may
in fact last longer and start earlier than when the class~II masers
are active.

\section{Conclusions}
\begin{enumerate}
\item The class~I methanol maser emission consists of a cluster of 6
  spots spread over an area of 30\arcsec{ }in extent. Five spots were
  detected in only the 84- and 95-GHz transitions (for two spots the
  84-GHz detection is marginal), while the sixth spot shows activity
  in all 12 observed transitions.
\item The three most southern maser spots show clear association with
  a jet-driven molecular outflow. Their velocities are close to that
  of the molecular core within which the jet is embedded. This fact
  supports the idea that the class~I masers reside in the interface
  regions of outflows.
\item Comparison with OH masers and infrared data reveals a potential
  discrepancy in the expected evolutionary state: the presence of the
  OH masers usually means that the source is evolved, but the infrared
  data suggest otherwise. The class~II methanol masers at 6.7~GHz
  are very widespread and are usually treated as an inevitable phenomenon at
  some stage of high-mass protostellar evolution (see section~\ref{evolution_section}).
  The lack of such a maser in this source (which is active in many
  class~I maser transitions) raises an additional question concerning
  its evolutionary state:
  whether this source is too young or too old to have a 6.7~GHz maser?
  We argue that both cases are possible because the class~I masers do
  not require an infrared source for pumping and are harder to destroy
  than the class~II masers, which are usually located close to the
  YSO. We, therefore, suggest that the evolutionary stage where the
  class~I masers are present may last longer than that with the
  class~II masers, although the exact status of this source is not
  clear at present.
\item We report the first interferometric observations of the rare
  9.9- and 104-GHz masers. The spectra for these transitions contain a
  very narrow spike ($<0.03$~\ks) which has a brightness temperature
  greater than $5.3\times10^7$ and $2.0\times10^4$~K at 9.9 and
  104~GHz, respectively.
\item High spectral resolution data leads us to suggest that the rest
  frequency for the $9_{-1}-8_{-2}$~E transition should be refined to
  9936.201$\pm$0.001.
\end{enumerate}

\section*{Acknowledgements}
We would like to thank Prof. Brian Boyle, the director of the
Australia Telescope, for granting an additional observing time for the
project on June 16. We also appreciate the efforts of an anonymous
referee, whose suggestions helped to improve the quality of this publication.
The Australia Telescope is funded by the
Commonwealth of Australia for operation as a National Facility managed
by CSIRO. SPE thanks the Australian Research Council for financial
support for this work.  This research has made use of NASA's
Astrophysics Data System Abstract Service.  This research has made use
of data products from the {\em Midcourse Space Experiment}.
Processing of the {\em MSX} data was funded by the Ballistic Missile
Defence Organization with additional support from the NASA Office of
Space Science.  This research has made use of data products from the
GLIMPSE survey, which is a legacy science program of the {\em Spitzer
Space Telescope}, funded by the National Aeronautics and Space
Administration.  The research has made use of the NASA/IPAC Infrared
Science Archive, which is operated by the Jet Propulsion Laboratory,
California Institute of Technology, under contract with the National
Aeronautics and Space Administration.


\bsp

\label{lastpage}


\begin{thebibliography}{99}

\bibitem[\protect\citeauthoryear{Avery \& Chiao}{1996}]{ave96}
Avery L.W., Chiao M., 1996, ApJ, 463, 642

\bibitem[\protect\citeauthoryear{Batrla \& Menten}{1988}]{bat88}
Batrla W., Menten K.M., 1988, ApJ, 329, L117

\bibitem[\protect\citeauthoryear{Benjamin et al.}{2003}]{ben03}
Benjamin R.A., Churchwell E., Babler B.L., Bania T.M.,
Clements D.P., Cohen M., Dickey J.M., Indebetouw, R., Jackson J.M.,
Kobulnicky H.A., Lazarian A., Marston A.P., Mathis J.S., Meade M.R.,
Seager S., Stolovy S.R., Watson C., Whitney B.A., Wolff M.J.,
Wolfire M.G., 2003, PASP, 115, 953

\bibitem[\protect\citeauthoryear{Brooks et al.}{2003}]{bro03}
Brooks K.J., Garay G., Mardones D., Bronfman L., 2003, ApJ, 594, L131

\bibitem[\protect\citeauthoryear{Brooks et al.}{2006}]{bro06}
Brooks K.J., Garay G., Voronkov M.A., Rodr\'\i guez L.F., 2006, ApJ, submitted

\bibitem[\protect\citeauthoryear{Caswell, Vaile \& Ellingsen}
{Caswell et al.}{1995a}]{cas95a}
Caswell J.L., Vaile R.A., Ellingsen S.P., 1995a, PASA, 12, 37

\bibitem[\protect\citeauthoryear{Caswell et al.}{1995b}]{cas95b}
Caswell J.L., Vaile R.A., Ellingsen S.P., Whiteoak J.B., Norris R.P.,
1995b, MNRAS, 272, 96

\bibitem[\protect\citeauthoryear{Cragg et al.}{1992}]{cra92}
Cragg D.M., Johns K.P., Godfrey P.D., Brown R.D., 1992, MNRAS, 259, 203

\bibitem[\protect\citeauthoryear{Codella \& Moscadelli}{2000}]{cod00}
Codella C., Moscadelli L., 2000, A\&A, 362, 723

\bibitem[\protect\citeauthoryear{Davis \& Smith}{1995}]{dav95}
Davis C.J., Smith M.D., 1995, 443, L41

\bibitem[\protect\citeauthoryear{Eisl\"offel, Smith \& Davis}
{Eisl\"offel et al.}{2000}]{eis00}
Eisl\"offel J., Smith M.D., Davis C.J., 2000, A\&A, 1147

\bibitem[\protect\citeauthoryear{Eisl\"offel et al.}{1996}]{eis96}
Eisl\"offel J., Smith M.D., Davis C.J., Ray T.P., 1996, AJ, 112, 2086

\bibitem[\protect\citeauthoryear{Ellingsen}{2005}]{ell05}
Ellingsen S.P., 2005, MNRAS, 359, 1498

\bibitem[\protect\citeauthoryear{Ellingsen}{2006}]{ell06}
Ellingsen S.P., 2006, ApJ, 638, 241


\bibitem[\protect\citeauthoryear{Forster \& Caswell}{1989}]{for89}
Forster J.R., Caswell J.L., 1989, A\&A, 213, 339

\bibitem[\protect\citeauthoryear{Garay et al.}{2002}]{gar02}
Garay G., Mardones D., Rodr\'\i guez L.F., Caselli P., Bourke T.L.,
2002, ApJ, 567, 980

\bibitem[\protect\citeauthoryear{Garay et al.}{2003}]{gar03}
Garay G., Brooks K.J., Mardones D., Norris R.P., 2003, ApJ, 587, 739

\bibitem[\protect\citeauthoryear{Garay et al.}{2006}]{gar06}
Garay G., Mardones D., Bronfman L., Brooks K.J., Rodr\'\i guez L.F.,
G\"usten R., Nyman L., Franco R., Moran J., 2006, A\&A, submitted

\bibitem[\protect\citeauthoryear{Gibb \& Davis}{1998}]{gib98}
Gibb A.G., Davis C.J., 1998, MNRAS, 298, 644

\bibitem[\protect\citeauthoryear{Goedhart, Gaylard \& van der Walt}
{Goedhart et al.}{2004}]{goe04}
Goedhart S., Gaylard M.J., van der Walt D.J., 2004, MNRAS, 355, 553

\bibitem[\protect\citeauthoryear{Habart et al.}{2005}]{hab05}
Habart E., Walmsley M., Verstraete L., Cazaux S., Maiolino R.,
Cox P., Boulanger F., Pineau Des For\^ets G., 2005, Sp. Sci. Rev., 119,
71

\bibitem[\protect\citeauthoryear{Hartquist et al.}{1995}]{har95}
Hartquist T.W., Menten K.M., Lepp S., Dalgarno A., 1995, MNRAS, 272, 184

\bibitem[\protect\citeauthoryear{Henriksen, Ptuskin \& Mirabel}
{Henriksen et al.}{1991}]{hen91}
Henriksen R.N., Ptuskin V.S., Mirabel I.F., 1991, A\&A, 248, 221

\bibitem[\protect\citeauthoryear{Kalenskii et al.}{2001}]{kal01}
Kalenskii S.V., Slysh V.I., Val'tts I.E., Winnberg A., Johansson L.E.,
Astron. Reports, 2001, 45, 26

\bibitem[\protect\citeauthoryear{Kalenskii et al.}{2005}]{kal06}
Kalenskii S.V., Promyslov V.G., Slysh V.I., Bergman P., Winnberg A.,
Astron. Reports, 2006, 50, 289 (astro-ph/0505225)

\bibitem[\protect\citeauthoryear{Kurtz, Hofner \& \'Alvarez}
{Kurtz et al.}{2004}]{kur04}
Kurtz S., Hofner P., \'Alvarez C.V., 2004, ApJS, 155, 149

\bibitem[\protect\citeauthoryear{Menten}{1996}]{men96}
Menten K.M., 1996, proceedings of IAU Symposium 178
(eds. E.F. van Dishoeck), 163

\bibitem[\protect\citeauthoryear{Menten et al.}{1986}]{men86}
Menten K.M., Walmsley C.M., Henkel C., Wilson T.L., 1986, A\&A, 157, 318

\bibitem[\protect\citeauthoryear{Mehringer \& Menten}{1996}]{meh96}
Mehringer D.M., Menten K.M., 1996, ApJ, 474, 346

\bibitem[\protect\citeauthoryear{Micono et al.}{1998}]{mic98}
Micono M., Davis C.J., Ray T.P., Eisl\"offel J., Shetrone M.D., 1998, ApJ,
494, L227

\bibitem[\protect\citeauthoryear{M\"uller, Menten \& M\"ader}
{M\"uller et al.}{2004}]{mul04}
M\"uller H.S.P., Menten K.M., M\"ader H., 2004, A\&A, 428, 1019

\bibitem[\protect\citeauthoryear{Nanda Kumar, Bachiller \& Davis}
{Nanda Kumar et al.}{2002}]{nan02}
Nanda Kumar M.S., Bachiller R., Davis C.J., 2002, ApJ, 576, 313

\bibitem[\protect\citeauthoryear{Pestalozzi et al.}{2002}]{pes02}
Pestalozzi M.R., Humphreys E.M.L., Booth R.S., 2002, A\&A, 384, L15

\bibitem[\protect\citeauthoryear{Phillips et al.}{1998}]{phi98}
Phillips C.J., Norris R.P., Ellingsen S.P., McCulloch P.M.,
1998, MNRAS, 300, 1131

\bibitem[\protect\citeauthoryear{Plambeck \& Menten}{1990}]{pla90}
Plambeck R.L., Menten K.M., 1990, ApJ, 364, 555

\bibitem[\protect\citeauthoryear{Pottage, Flower \& Davis}{Pottage et al.}{2004}]{pot04}
Pottage J.T., Flower D.R., Davis S.L., 2004, J. Phys. B., 37, 165

\bibitem[\protect\citeauthoryear{Reipurth \& Bally}{2001}]{rei01}
Reipurth B., Bally J., 2001, ARA\&A, 39, 403

\bibitem[\protect\citeauthoryear{Rodr\'\i guez et al.}{2005}]{rod05}
Rodr\'\i guez L.F., Garay G., Brooks K.J., Mardones D., 2005, ApJ, 626, 953

\bibitem[\protect\citeauthoryear{Sandford \& Allamandola}{1993}]{san93}
Sandford S.A., Allamandola L.J., 1993, ApJ, 417, 815

\bibitem[\protect\citeauthoryear{Slysh et al.}{1994}]{sly94}
Slysh V.I., Kalenskii S.V., Val'tts I.E., Otrupcek R., 1994, MNRAS, 268,
464

\bibitem[\protect\citeauthoryear{Smith \& Brand}{1990}]{smi90}
Smith M.D., Brand P.W.J.L., 1990, MNRAS, 245, 108

\bibitem[\protect\citeauthoryear{Salii, Sobolev \& Kalinina}
{Salii et al.}{2002}]{sal02}
Salii S.V., Sobolev A.M., Kalinina N.D., 2002, Astron. Rep., 46, 955

\bibitem[\protect\citeauthoryear{Sobolev}{1992}]{sob92}
Sobolev A.M., 1992, Soviet Astron., 36, 590

\bibitem[\protect\citeauthoryear{Sobolev et al.}{2005}]{sob05}
Sobolev A.M., Ostrovskii A.B., Kirsanova M.S., Shelemei O.V.,
Voronkov M.A., Malyshev A.V., 2005, proceedings of IAU Symposium 227
(eds. E.Churchwell, P.Conti and M.Felli), 174 (astro-ph/0601260)

\bibitem[\protect\citeauthoryear{Sutton et al.}{2004}]{sut04}
Sutton E.C., Sobolev A.M., Salii S.V., Malyshev A.V.,
Ostrovskii A.B., Zinchenko I.I., 2004, ApJ, 609, 231

\bibitem[\protect\citeauthoryear{Szymczak, Pillai \& Menten}
{Szymczak et al.}{2005}]{szy05}
Szymczak M., Pillai T., Menten K.M., 2005, A\&A, 434, 613

\bibitem[\protect\citeauthoryear{Val'tts et al.}{2000}]{val00}
Val'tts I.E., Ellingsen S.P., Slysh V.I., Kalenskii S.V.,
Otrupcek R., Larionov G.M., 2000, MNRAS, 317, 315

\bibitem[\protect\citeauthoryear{van der Walt}{2005}]{vdw05}
van der Walt D.J., 2005, MNRAS, 360, 153

\bibitem[\protect\citeauthoryear{Voronkov}{1999}]{vor99}
Voronkov M.A., 1999, Astron. Lett., 25, 149 (astro-ph/0008476)

\bibitem[\protect\citeauthoryear{Voronkov et al.}{2005a}]{vor05a}
Voronkov M.A., Sobolev A.M., Ellingsen S.P., Ostrovskii A.B., Alakoz A.V.,
2005a, ApSS, 295, 217 (astro-ph/0407275)

\bibitem[\protect\citeauthoryear{Voronkov et al.}{2005b}]{vor05b}
Voronkov M.A., Sobolev A.M., Ellingsen S.P., Ostrovskii A.B.,
2005b, MNRAS, 362, 995 (astro-ph/0507048)

\bibitem[\protect\citeauthoryear{Walsh et al.}{1998}]{wal98}
Walsh A.J., Burton M.G., Hyland A.R., Robinson G., 1998, MNRAS, 301, 640

\bibitem[\protect\citeauthoryear{Walsh, Lee \& Burton}{2002}]{wal02}
Walsh A.J., Lee J.-K., Burton M.G., 2002, MNRAS, 329, 475

\bibitem[\protect\citeauthoryear{Walsh et al.}{2003}]{wal03}
Walsh A.J., Macdonald G.H., Alvey N.D.S., Burton M.G., Lee J.-K.,
2003, A\&A, 410, 597

\bibitem[\protect\citeauthoryear{Wiesemeyer, Thum \& Walmsley}
{Wiesemeyer et al.}{2004}]{wie04}
Wiesemeyer H., Thum C., Walmsley C.M., 2004, A\&A, 428, 479

\end{thebibliography}
\end{document}